# An alternative method of image simulation in high resolution transmission electron microscopy

Usha Bhat,[1,2] and Ranjan Datta[1,2,*]

[1,2]*International Center for Materials Science, Chemistry and Physics of Materials Unit, JNCASR, Bangalore 560064.*

**Abstract**

An alternative approach to the image simulation in high resolution transmission electron microscopy (HRTEM) is introduced after comparative analysis of the existing image simulation methods. The alternative method is based on considering the atom center as an electrostatic interferometer akin to the conventional off-axis electron biprism within few nanometers of focus variation. Simulation results are compared with the experimental images of 2D materials of MoS$_2$, BN recorded under the optimum combination of third order spherical aberration $C_s = -35$ μm and defocus $\Delta f = 1, 4,$ and 8 nm and are found to be in good agreement.

Corresponding author e-mail: *ranjan@jncasr.ac.in*

## 1. Introduction

Imaging of any object both in transmission and reflection geometry is generally carried out by detecting the scattered (incoherent) and diffracted (coherent) radiation on a recording device e.g., a camera placed at different reference planes away from the object on the optic axis. Image is the replica of the object and not the object itself and the information about the object is carried to the detector through the complex wave function.[1–5] Maximum spatial details that can be obtained is limited by the diffraction and the microscope performance. The entire topic of quantitative HRTEM falls into two broad categories; (*i*) object exit wave (OEW) reconstruction, to retrieve the missing phase information from the recorded image, and (*ii*) image simulation to interpret the OEW with the object structure. Various schemes are available for the reconstruction of OEW function to recover the phase related to the object potential and the crystallography and will not be discussed here, for details see Ref. [6–8]. There are several aspects in HRTEM image simulation that need to be considered e.g., probe electron, interaction between the fast electron and the specimen potential, lens action and characteristics of the recording device.[5,9] The probe illumination is typically a plane wave of electron with relativistic energy in the range of 100-300 kV ($\lambda = 1.97\ pm$ at 300 kV). The amplitude $A(x, y)$ and the phase $\phi(x, y)$ of the OEW function of the form $\psi = Ae^{i\phi}$ extracted from the recorded intensity pattern are used to interpret the potential information of atoms in solid. The information on object potential can be used to extract wide range of information such as number of atoms or thickness along the beam propagation direction, identification of atoms, valence electron sharing between the atoms etc.[7,10–14] However, the change in phase ($\phi$) of the probe electron wave after interaction with the specimen potential and resulting modulation in intensity pattern has been treated in fundamentally distinctive ways e.g., (*i*) transmission function based on weak phase object approximation (WPOA) along with Zernike type π/2 or λ/4 phase plate equivalent to phase contrast transfer function (PCTF) to account for the lens aberration, where the phase change is incorporated in terms of change in magnitude of momentum vector $k$ of the probe electron due to specimen potential,[4,5] (*ii*) phase change according to scattering amplitude in terms of atom scattering $f(k)$ and structure factor $F(g)$ along with PCTF,[9,15,16] and (*iii*) self-interference in HRTEM and holographic fringe shift in off-axis electron holography where a phase term (±φ) is added inside the trigonometric functions with respect to the reference phase.[17,18] Kindly note that the phase change due to aberration through PCTF is not added with the object wave rather it is applied as frequency filter and point spread function (*psf*) in the diffraction and image planes, respectively.

Therefore, in the present manuscript, at first a comparative analysis is provided on the various existing methods for simulating the image of the atom. It is shown that the different ways of considering phase change in the probe electron wave function due to atomic potential results in different magnitude and intensity pattern for the same atom. Subsequently, an alternative method is introduced where the geometry of interference based on direction of momentum vector is emphasized. The method is based on atomic potential center as an interferometer akin to the electron biprism within short range of focus variation (<10 nm) from the reference Gaussian image plane and has resemblance with the Abbe's picture of diffraction. In this alternative method, the aspect of phase change has been treated like off-axis electron holography considering the wave interference at an angle and its analogy with other approaches can be understood with the help of interference geometry and associated momentum vector direction. Addressing the effect of large defocus together with various coherent aberration envelopes, and thickness on the rich variation in phase and image intensity patterns by the alternative procedure will be part of a separate discussion as this requires a departure from the traditional viewpoint due to unique experimental observation overlooked in the past. Simulation results are compared with the experimental image of 2D materials of MoS$_2$ and BN recorded under specific settings of third order spherical aberration $C_s = -35$ μm and defocus $\Delta f = 1, 4,$ and $8$ nm and are found to be in good agreement.

## 2. Coherent image formation at near and far field

To begin with, a brief discussion is provided on the existing methodologies involved in the image simulation of atoms. This has analogy with the slit diffraction pattern in light optics both at near and far field regimes. A typical Fresnel and Fraunhofer diffraction regimes and corresponding patterns are shown in Fig.1. In light optics, there exist analytical formula derived from the Fraunhofer integral and Fourier transformation-based method to evaluate the far field diffraction pattern [sec. S2]. The analytical formula embodies various parameters e.g., wavelength (λ) of the illumination, scattering angle (θ), dimension and periodicity of the slits. The approach is based on the physical picture of path difference and associated constructive and destructive interference between waves having same momentum vector direction described by the plane wave front from a pair of spatial points at the slit opening that ensures the distribution of intensity transfer along different scattering angles. This is different compared to diffraction geometry involved in Fresnel zone construction for image formation at near filed [Fig. 1(c)]. In case of Fresnel diffraction, it is the angular correlation between the wave vectors pointing at different direction lying on the surface of a sphere and the phase difference between various wave vectors is acquired due to path difference of waves while converging to a point with respect to outward curvature of the spherical wave front (more precisely parabolic wave front in case of Fresnel regime) [Fig. S2 &S3]. This is the basis of Fresnel zone construction. Kindly note that the there is no path difference between various wave vectors with respect to the emitting point as the radius is same for a sphere, but not so for a parabolic and plane wave front geometry. For comparison with the off-axis electron holographic interference geometry and present alternate method, see sec. 2.3.1.

On the other hand, in Fourier transformation (FT)-based approach, the far field image of the object is the modulus of the FT (abs-FT) of the object function. However, the phase angle calculated for Fourier waves or equivalently Abbe waves (see Abbe's two step imaging process in sec. S2.3) corresponding to each frequency do not have the information on the scattering angle. Thus, the frequency of Fourier waves needs to be calibrated either with respect to the scattering angle obtained through analytical method or from an experiment using a standard sample with known lattice parameter [Fig. S6].

Now in the context of diffraction from atomic potential, the Fraunhofer pattern is the atom scattering amplitude $f(k)$ and the structure factor $F_g(k)$ for isolated and periodic atoms, respectively that depends on the strength and crystallography of the scattering potential in the object space. Figure 2 shows example scattering distribution of isolated B and Mo atoms and in monolayer hexagonal periodic BN and $MoS_2$ lattice. In the image plane, the transmitted wave function or object exit wave (OEW) function can be derived in different ways, e.g., weak phase object approximation (WPOA) according to Zernike, Schrodinger integral equation and evaluating the Fraunhofer integral and are discussed next.[9,16]

**2.1. Image simulation based on Zernike phase object and WPOA**

This has origin in Zernike phase contrast theory where for pure phase or weakly scattering object, the object transmission function is represented by a complex function of the form $F(x) = e^{i\phi(x)}$ or $\sim 1 + i\phi(x)$ for small $\phi$, where $\phi$ is a real phase function corresponding to the discrete or periodic transparent object. This is known as weak phase object approximation (WPOA).[1] In fact, the object function is real, it is the replica of the object in the form of object wave that carries the information encoded into its phase and amplitude and is similar in function to the Fourier or Abbe waves and holographic direct and twin image wave components. The effect of Zernike phase plate modifies the intensity of the object wave that depends linearly on the object phase according to $I(x) = 1 \pm 2\phi(x)$.

WPOA is a straightforward and widely applied approach to simulate the HRTEM images of thin samples. In HRTEM, WPOA describes the phase shift of probe electron wave due to object electrostatic potential projected along the beam propagation direction, and the transmission function has the following expression after invoking WPOA i.e., ignoring the terms with $\sigma^2$ and higher order,

$$t(x) = 1 - i\sigma V_t(x, y) \qquad (1)$$

Where, $V_t(x, y)$ is the projected specimen potential and $\sigma = \frac{2\pi me\lambda}{h^2}$ is the interaction constant. The approximation in Eq. 1 is essential to retain the object information in the image plane. The accompanied transmitted wave function is derived considering the change in magnitude of electron wavelength from $\lambda$ to $\lambda'$ due to attractive positive potential ($V_s$) of the specimen. Refraction of electron through atomic potential is associated with the change in momentum vector direction as well and is addressed in the description of alternative method (sec. 2.3). $\lambda'$ can be an average change corresponding to a mean inner potential (MIP) for a given spatial extent or as a function of spatial position from the center of the atom at medium and atomic resolution, respectively. At medium resolution, for average projected potential $V_t$, the transmitted wave function of the electron within kinematical scattering in 1D is given by

$$\psi_t(x) \sim t(x) \exp(2\pi i k_z z) \qquad (2)$$

Eq. 2 is equivalent to the reading component of Gabor's in-line holography (sec. S1). The plane wave component in Eq. 2 contributes to the background as DC component and poses difficulty in in-line holography along with twin image components.

The image intensity after considering for the lens effect is given by [4,5]

$$I(x,y) = \psi_i(x,y)\psi_i^*(x,y) \approx 1 + 2\sigma\phi_p(-x,y) * \mathcal{F}\{sin\chi(u,v)P(u,v)\} \qquad (3)$$

Fig. 3 shows the image intensity calculated by using Eq. 3 with and without considering the lens response for isolated Mo, S, N, and B atoms.[7] Not considering lens response is similar to Zernike like phase transfer and as the potential function is asymptotic, peak value will remain undefined with a background value of 1 [Fig. 3(b)]. Considering aberration through optimum PCTF ($C_s = -35\ \mu m$ and $\Delta f = 8\ nm$), peak values (and FWHM) of ~ 22000 (0.25 Å) and 3500 (0.25 Å) are obtained for Mo and B atoms, respectively [Fig. 3(a)]. This gives a peak intensity ratio of ~ 6.2. Peak intensity increases linearly with atomic number [Fig S14]. The trend is in contrast with experimental observation where changes in peak intensity in the first decimal place is observed with the atomic number [Table 1]. The high peak value according to Eq. 3 is due to convolution procedure and cannot be normalized individually as the image without PCTF is not known [Fig. S9]. Periodicity can be extended through the lattice vectors in the image plane for the images of $MoS_2$ and BN lattice.

As already mentioned, that the phase shift due to aberration cannot be added in the trigonometric function in diffraction plane as that shifts the wave amplitude, rather it is used as coherent envelope function or frequency filter in the diffraction plane and as point spread function (*psf*) in real space. According to Eq. S28, the PCTF gives weight to the magnitude of *psf* and aperture function in the form of Bessel function equivalent to Abbe's theory that sets the resolution in terms of full width at half maximum (FWHM) in the final image.[5] Scherzer phase transfer will have maximum weight for optimum value of spherical aberration and defocus that depends on the integration value of PCTF over the band pass limits of spatial frequency. The weight of specimen potential in Eq. 3 has the similar effect as atom scattering factor in Eq. 6. As the convolution procedure changes the magnitude of the resultant function significantly [Fig. 3(a)], a flux balance approach is helpful to observe the qualitative decrease in intensity and increase in FWHM due to aberration compared to ideal image free from any aberration (Fig. S9). The method described in sec. 2.3 can be used as a reference atom image without any lens response to employ quantitative flux balance after application of PCTF in order to compare the effects with change in $C_S$, $\Delta f$ and resolution.

At medium resolution, it is the mean inner potential (MIP) extracted from the mean phase shift has been utilized so far. However, in case of single atom at sub-atomic resolution, variation of potential around the atom is important and can be used for various other studies.[7] Moreover, the square amplitude of transmitted wave function will be unity unless series approximation/weak phase object approximation is conjured and no information on phase can be obtained [sec. S2.6.1] unless a reference wave is used in the mathematical expression according to in-line or off-axis holographic geometry.[6,17] In fact, the same transmitted wave function for HRTEM can also be written considering self-interference between plane wave and the scattered wave that is equivalent to writing component of Gabor's in-line holography instead of reading as given by Eq. 2 [Eq. S4]. However, the above description based on WPOA does not have any information on the geometry of interference in terms of momentum vector directions. Instead, considering the change in momentum vector direction due to interaction with the object potential and ensuing interference effect draw a clear comparison between the various pictures (sec. S2.1 and S2.2).

### 2.2. Image simulation based on atom scattering factor

Another approach of image simulation incorporates atom scattering factor directly instead of specimen potential. The transmitted wave function in this case can be derived from the Schrödinger integral equation and has the following form[9,19]

$$\psi_t(x) = \exp(2\pi i k_z z) + f_e(q) \frac{\exp(2\pi i q.r)}{r} \qquad (4)$$

Where, $q = k - k_0$, and $f_e(q)$ is the atom scattering factor and is defined by,

$$f(q) = -\frac{m}{2\pi \hbar^2} \int V(r') e^{-2\pi i q.r} d^3r \qquad (5)$$

Which is the FT of the scattering potential. The solution of wave function based on differential form of Schrödinger equation has the form of a plane wave. On the other hand, integral form gives solution of spherical waves along with amplitude factor as atom scattering factor. The equivalence between the two solutions can be perceived in terms of envelop of all the spherical waves from many adjacent scattering centers will eventually result in a plane wave front. The picture is akin to Huygens's construction that a plane wave front is the envelop of many forward scattered spherical wavelets and equivalent to First Born approximation. This result is used along with the scattering factor as derived by Moliere to calculate the image of isolated atoms by using the following equation.[9,16]

$$g(x) = \left| 1 + 2\pi i \int_0^{k_{max}} f_e(k) \exp[-i\chi(k)] J_0(2\pi k r) k dk \right|^2 \qquad (6)$$

Where, $f_e(k)$ is the electron scattering factor in the Moliere approximation which has the advantage over Born scattering factor due to the presence of imaginary component. $\chi(k)$ is the aberration function, $k_{max} = \alpha_{max}/\lambda$ (rad Å$^{-1}$) is the maximum spatial frequency allowed by the objective aperture and $J_0(x)$ is the Bessel function of order zero. The intensity expression used by Scherzer has the similar form as given in Eq. 6 which is derived based on Fraunhofer approximation.[16] Fig. 4 shows the example intensity pattern for four different atoms considering PCTF and complete CTF with $Cs = 35$ μm, $\Delta f = 8$ nm.

The phase contrast image calculated using above expression varies weakly with atomic number and the peak phase shift $\varphi_{max}(rad)$ follows ~ $Z^{0.6} - Z^{0.7}$, where $Z$ is the atomic number.[9,20] Though the trend can be complicated depending on the valence electron filling and for specific atoms with higher $Z$ can have smaller contrast compared to atoms with lower Z next to each other in the periodic table.[9,17,21] The peak intensity is almost the same irrespective of the atomic number and changes only in the second decimal place. Fig. S14 and Table S1 summarized the peak intensity and FWHM values calculated using Eq. 3 (WPOA) and Eq. 6 (atom scattering factor) for Mo, S, B, N, Zn and O atoms. One can notice that the difference in peak intensity and FWHM maximum calculated by two different methods are markedly different. The intensity values calculated based on Eq. 6 shows frivolous dependence considering only PCTF irrespective of atom number. However, the peak values are much smaller and FWHM are higher by a factor of two, respectively calculated by Eq. 6 and Eq. 3, and the difference will remain even after the flux balance.

### 2.3. Atom as an interferometer

In this section, the alternative method based on the concept of atom as electrostatic charge center and its action as an interferometer on the simulation of atom image is described. The size of the atomic nucleus is extremely small (~1.6-15 fm) compared to the overall size of the atom with electron clouds (~ 0.1-0.5 nm). Therefore, the nucleus can safely be considered as a source of positive point charge (+$Ze$) with associated Coulomb potential that decay inversely away from the charge center [Fig. 5(a)]. The electron clouds surrounding the nucleus only screens the radially symmetric positive Coulomb potential. Moreover, the surrounding electrons clouds scatter the probe electron inelastically.[22] The probability of inelastic events for the fast probe electron is negligible compared to the dominant elastic events for samples having thickness less than the extinction length and at short exposure time typically 1-2 sec. This screened positive potential is attractive to the negatively charged probe electron wave while passing by the nucleus. The force experienced by the traversing electron will depend on the distance away from the nucleus. The calculated screened projected potential of Mo atom following the Hartree-Fock model[9] and associated attractive force in terms of bending angle is shown in Fig. 5(b). Within this picture the projected potential of an atom can be considered as an electrostatic circular prism similar to the experimental off-axis electron holography biprism except attractive force acting from all directions with varying strength as a function of radial distance around it. This implies that the atom bends electron trajectory along $\theta = 0 - 2\pi$ azimuthal direction around it where momentum vectors lying on a conical surface representing scattering angles within same order of magnitude whereas a cylindrical biprism does the same but along single pair of momentum direction across a mirror plane [Fig. 6]. Therefore, for atomic case it is necessary to describe the interference pattern from a different geometrical perspective than typical unidirectional electrostatic bi prism. The picture is founded on the classical concept of wave optics but differs from the concept of channeling of quantum mechanical Bloch waves of electron through the crystal lattice. The correspondence between the two different pictures and the evolution of far field diffraction pattern will be part of a separate discussion. In the following sub-sections, at first the method based on off-axis electron holography is presented briefly to introduce on the formation of image contrast and then extending the principle to 'the atom as electrostatic charge center' to simulate the image of the atom.

### 2.3.1. Image contrast in off-axis electron holography with electron biprism

The formation of electron interference pattern and resulting contrast in off-axis electron holography is emphasized here. The basic principle is based on single electron wave interference and the expression corresponding to HRTEM is given in Eq. S4 [sec. S1]. The intensity pattern of the hologram is given by Eq. 7, details on the off-axis electron holography methods and practices can be found in Ref.[17,18].

$$I(x,y) = I(x) = a_1^2 + a_2^2 + 2a_1 a_2 \cos(2\pi q_c x + \Delta\phi) \qquad (7)$$

Where, $q_c = 2k_x$ is carrier spatial frequency of the hologram. $a_1$ and $a_2$ are the amplitudes of waves undergoing interference at an inclination angle due to action of biprism. $x$ is the spatial coordinate variable of the interference pattern and $\Delta\phi$ is the difference in phase between the two interfering partial waves, typically acquired by one of the two waves due to object potential. $\Delta\phi = 0$ for the vacuum wave. Information on $\Delta\phi$ appears as a small deviation in hologram fringe. The maxima and minima of the intensity pattern can be determined from Eq. 7 and remains the same throughout the interference field. The carrier frequency $q_c$ can be considered as carrying the information about the field strength and associated potential of the biprism. Field strength, potential and carrier frequency can be empirically put together in a functional form. Now, it is important to note that the total flux or energy of the interfering waves must be preserved on the resulting interference field. This will become essential while calculating the image intensity and distinguishing contrast between the various atoms upon extending the above principle for atom as charge center.

The $q_c$ of the hologram that represents the fringe wavelength depends on the angle of superposition for a given wavelength. Higher the angle of superposition larger will be the carrier frequency resulting in finer fringe spacing due to horizontal component of the wave vector is larger at higher inclination angle. This can be thought of as equivalent to the Abbe's picture where wave vector is larger or higher frequency at higher scattering angle [sec. S2.3]. Thus, the two pictures may be unified and said that that higher the inclination angle, smaller will be the fringe spacing or inter feature distance resulting in better spatial resolution. This was the Abbe's hypothesis in describing the diffraction limited imaging, where restricting the higher frequency by numerical aperture coming at a higher scattering angle limits the spatial resolution. However, in case of off-axis electron holography the geometry of interference involves only pair of momentum vector directions across a mirror plane passing through the biprism compared to set of momentum vectors directions lying on the surface of cones with continuously varying slant length in case of Abbe's geometry.

### 2.3.2. Image contrast due to atom charge center equivalent to biprism

Now, the image formation by interference of waves due to atom charge center similar to 1D electrostatic bi prism is described here. Considering radially symmetric atomic potential the intensity pattern can be calculated following Eq. 7 after incorporating wave interference effect from a given radial zone of extent $\Delta r$. The zones described here are similar to binary type Fresnel zone plates with multiple foci which depends on the scattering angle with different order of magnitude. Calculating interference pattern along all azimuthal inclination angle for the peripheral zone area requires two additional considerations compared to the unidirectional interference pattern. The first consideration is that the wave flux will depend linearly on the perimeter $2\pi r$ which is a function of radial distance $r$ from the center of the atom [Eq. 8 & sec. S2.9]. Larger the perimeter or the zone area [ $\pi(r_2^2 - r_1^2)$] away from the atom center, higher will be the flux of the wave approaching for the interference. The relative intensity contribution at the center of the pattern from different rims belongs to the same spatial coherent zone is scaled with $2\pi r$, where $r$ is the radial distance from the center of the atom [Fig. 6, Eq. 8].

$$I_{rad\ int}(r) = a_1^2 + a_2^2 + 2a_1 a_2 \cos(2\pi q_c r) * 2\pi(r_{max} - r) \tag{8}$$

The second consideration is that of flux balancing between the flux of wave at the plane of the atom as given by the coherent rim area $\pi(r_2^2 - r_1^2) * I$, where $I$ is the intensity at a given pixel point within this zone, and resulting interference field over a circular area around the optic axis as given by $\pi r^2 \times I_{rad\ int}$ (1$^{st}$ law of thermodynamics).

$$\int_{-drr}^{drr} I(r)\, dr = \pi(r_2^2 - r_1^2) \times 1 \tag{9}$$

Where, $drr = (r_2 - r_1)/2$ and $r = \sqrt{x^2 + y^2}$. $I = 1$ is the minimum intensity count on a pixel of size of 1 pm considered for the present calculation.

The resulting interference pattern is different compared to the unidirectional interference geometry. The pattern is radially symmetric with peak intensity at the origin and falls in amplitude away from the center unlike unidirectional pattern where average intensity remains the same throughout the field of superposition [Fig. 6]. There is another distinct difference between the two interference processes in terms of illumination geometry and the effective range of electrostatic field maintaining the spatial coherency which pulls the waves from opposite directions to interfere. In case of off-axis holography, it is well known that the contrast of hologram improves significantly if elliptical probe is used instead of a round probe [Fig. 7]. This is due to superior spatial coherency associated with the elliptical shape compared to round shape.[17] However, the spatial coherency of the wave can be correlated with the order of electric field magnitude for a given spatial extent and equivalent order of bending angle due to Coulomb attraction with respect to the charge center. This is in contrast with the earlier consideration of effective decrease in angular distribution of illumination aperture associated with the elliptical probe and high spatial coherency.[23] For elliptical spread of beam, most of the spatial extent of wave will approach from electric field regions away from the biprism center and will have bending angle almost within the same order of magnitude [Fig. 7]. This is the regions of high spatial coherency resulting in high contrast compared to regions close to bi-prism charge center where field strength and associated inclination angle varies strongly with the spatial distance. The situation has the analogue with high energy electron diffraction at small angle vs large angle scattering and their relationship with the spatial coherency. At small angle, coherency in elastic scattering is well preserved and gives rise to crystal Bragg diffraction peaks whereas at larger angles, scattering becomes more and more incoherent i.e., Rutherford type scattering. However, for atom case, there is no advantage for shaping the probe elliptical as potential falls much more rapidly, presence of neighboring atoms in case of crystal and short range of potential compared to bi-prism case. In case of atom, the potential varies strongly close to the nucleus and slowly beyond certain distance away. Therefore, there will be different contribution from various radial zones with specific rim width defined by the angle of superposition within the same order of magnitude to the overall pattern that can be controlled by the lens focus [Fig. 8]. The lens focus will effectively increase the width of interference field for the coherent zone. Therefore, we have divided the entire spatial range surrounding the atom starting from 1 pm to suitable outer range in terms of various zones with respect to example 1, 4 and 8 nm focus step (experimental through focus condition) and calculated intensity pattern for different zones separately and then summing up the contributions [Fig. 8 & 9]. The contributions to the overall intensity pattern from different zones are incoherent and can be thought of as either binary Fresnel zone plate already mentioned or annular Airy apertures giving rise to Airy pattern at specific focal length. Each zone is acting as a filter to specific momentum vector components similar to the PCTF as spatial frequency filter. The spatial extent of zone increases away from the atom due to decreasing potential and associated scattering angle. For example, the outer spatial extent is 10.31 and 4.19 pm corresponding to 1 nm focal length on the optic axis for Mo and B atoms, respectively and extent of the outer zone contributing to the coherent interference field increases with increasing focus (Fig. 8, S15, Table S2). Now depending on the focus settings various zones will contribute to intensity pattern (at focus) and background (away from focus) differently. For a given intermediate focus setting, the focus length smaller and larger than that will contribute to the background differently. The zones closer to atom center would have formed interference pattern and propagate with the information along the original direction and various diffracted direction. Diffracted interference patterns will appear as displaced image with intensity contribution very small compared to direct pattern almost an order of magnitude smaller which is considered to be similar in terms of relative total intensity between CB and SBs in off-axis and between direct beam and diffracted beams in HRTEM. Fig. 9 is showing schematically various contributions from different zones for example Mo atom case for 1 nm. Moreover, the flux contributions will be different from different zones and will be proportional to the annular zone area. As already mentioned, that the area and corresponding flux will be higher for the zones away from the nucleus. The calculated image of Mo atom at three different focus values are shown in Fig. 10 (a). The image of lighter atom B for similar focus settings is given in Fig. S16. Kindly note that, with increasing focus the zones with large radial distance will contribute and modify the peak value of the image. The peak values are smaller for B compared to Mo case as the extent of outer zone is reduced due to smaller magnitude to potential field. Fig. 10(a) is the graph of peak image intensity vs atomic number for three different focus values. One can see that with larger focus value the peak intensity decreases slightly. The peak intensity has a dependence of $\sim AZ^B$, where $A$ is a fitting constant and $Z$ is the atomic number. Exponent B changes from 0.5 to 0.26 for change in focal length from 1 to 8 nm.

However, the peak intensity values decrease significantly after considering image aberration (sec. 2.3.3 and Fig. 10(b)).

The relationship between the phase of the OEW and resulting magnitude of intensity can be understood as follows. Here the meaning of phase change is the change in momentum vector direction, which is similar to Fresnel, Fraunhofer, off-axis electron holography type of interference geometry where wave interference is involved at an angle and different than the WPOA type phase shift where no information on geometry of interference is available. Radial interference geometry modifies the unidirectional straight electron interference fringe to a radial symmetric pattern with a peak intensity at the center of the pattern. Whereas the intensity pattern oscillates periodically with the same magnitude in case of unidirectional interference geometry across a mirror plane in off-axis electron holography. Now this phase term appears as carrier frequency $Q$ in off-axis electron holography and in the present case alters the peak intensity of the radial pattern depending on field strength and extent of a given radial zone. The field strength and corresponding potential information is the object information that can be interpreted and used to identify atoms. In case of WPOA, series approximation and PCTF contrast is required to observe intensity pattern due to object potential but in the present case only defocus is sufficient. Now, in practical HRTEM, the atom contrast never goes to zero even if PCTF is zero. Traditionally, this is explained based on amplitude contrast due to ACTF which is the higher order contribution from interaction constant $\sigma$ according to Eq. 3 and is attributed as non-linear imaging condition. However, the presence of contrast even if defocus and corresponding PCTF are zero, can only be explained by the present method in terms of radial interference originating from sectors of potential very close to the nucleus. Therefore, in the present method as the meaning of phase is different and different incoherent zones will have different phase change in terms of $Q$, thus there will be more than one wave functions superimposed on top of each other incoherently and it will be appropriate to interpret intensity directly to the object information. Alternative proposal on image reconstruction based on intensity directly instead of wavefunctions is discussed elsewhere.

### 2.3.3. Effect of aberration and defocus on the image series

The effect of third order spherical aberration ($C_s$) and defocus ($\Delta f$) are now considered to modify the image contrast further. The image contrast calculated based on the alternative method without consideration of lens transfer does not match with the experimental images as a function of atomic number recorded under the particular settings of imaging condition. Generally, a suitable combination of $C_s$ and $\Delta f$ are used to transfer maximum phase contrast for a given band width of spatial frequency with best possible point resolution as given by the first zero crossing of phase contrast transfer function (PCTF) [Fig. S18]. A point resolution of better than 1 Å can be achieved in an aberration corrected microscope for a combination of $C_s = -35\ \mu m$ and $\Delta f = 8\ nm$.[24] As already mentioned that the effect of aberration phase shift can be considered either in diffraction plane or image plane depending on the requirement. In the image plane the aberration figure due to spherical aberration is given by a disk of radius $r_s$ at the Gaussian image plane [sec. S.2.10]. The effect of $C_s$ and $\Delta f$ act in opposite directions which depends on the third power and linearly with the scattering angle, respectively. The resulting effect is to either increase or decrease the full width at half maxima (FWHM) of the point spread function (*psf*) and impairs or improves the resolution of the imaging process, respectively. A typical angular deviation from the ideal ray path and corresponding aberration figure can be correlated for a given combination of $C_s$ and $\Delta f$ [Fig. S17 and Table S3]. Now depending on the electrostatic zones and associated scattering angle, the diffraction limited information will be blurred further for a given settings of $C_s$ and $\Delta f$.

Contributions of peak intensity from various incoherent zones can be found in Table S2. The peak intensity will be reduced due to aberration blurring for zones which have scattering angle more than what can be adjusted through PCTF. The effect of Peak image intensity and contrast difference between Mo and B atoms after considering $C_s = -35$ μm and $\Delta f = 1, 4$ and $8$ nm is shown in Fig. 10 (b) and listed in Table 1. The modified peak intensity has been calculated based on the flux balance approach considering the ratio between the original interference field width to the final blurred area. Kindly note that the simulated peak intensity values at focus close to zero remain little over reference background 1 irrespective of atomic numbers. This corresponds well with the experimental observation [Table 1].

Now, one can notice that after considering aberration blurring due to *psf* corresponding to each zone and scattering angle, the peak intensity falls significantly compared to ideal lens case [Table 1 and Fig. 10(b)]. From the experimental image series recorded under similar conditions as simulation parameters, the peak intensity to reference vacuum ratio is ~ 1.23 (± 0.04) and 1.05 (± 0.02) for Mo and B atoms, respectively after normalizing with respect to vacuum intensity. For example, we have obtained a peak value of 1.19 and 1.09 for Mo and B atoms from calculation for a focus setting of 4 nm (Table 1). The peak intensity does not vary significantly in the experimental image for small variation of focus and a good matching is obtained with the simulated image after considering aberration blurring of intensity from different incoherent zones. Any deviation can be explained in terms of slight deviation in scattering angle and associated extent of zones, focus value, consideration of aberration figure while simulating the absolute intensity of atoms. The close agreement between experimental image and simulation results is in contrast with earlier prediction based on Stobbs factor where a difference in contrast by a factor of 4 was reported .[25]

3. **Conclusions**

In conclusion, an alternative approach for simulating atom image in HRTEM is presented. The method is based on atom as electrostatic charge center inducing interference of wave along radial direction and an extension of the conventional off-axis interference geometry with unidirectional electron biprism. The simulated results corresponding to the image intensity of various atoms are in close agreement with the experimental images of Mo and B atoms.

**Acknowledgement**

The authors sincerely thank ICMS and JNCASR for the financial support. Prof. Ranjan Datta would like to give a special thanks to Deutsche Forschungsgemeinschaft (DFG) for providing funding for an illuminating and exceptional research visit in TU-Berlin in the year 2016.

**Table 1:** Peak intensity after considering aberration blurring for $C_s = -35\ \mu m$ and $\Delta f = +1, +4,$ and $+8$ nm in comparison with experimental image recorded under similar condition

| Defocus (nm) | Peak Intensity Mo (exp±0.04) | Peak Intensity B (exp±0.02) | Peak Intensity Mo (sim) | Peak Intensity B (sim) |
|---|---|---|---|---|
| **+1** | 1.08 | 1.04 | 1.03 | 1.02 |
| **+4** | 1.23 | 1.05 | 1.19 | 1.09 |
| **+8** | 1.24 | 1.05 | 1.25 | 1.11 |

**Figure:**

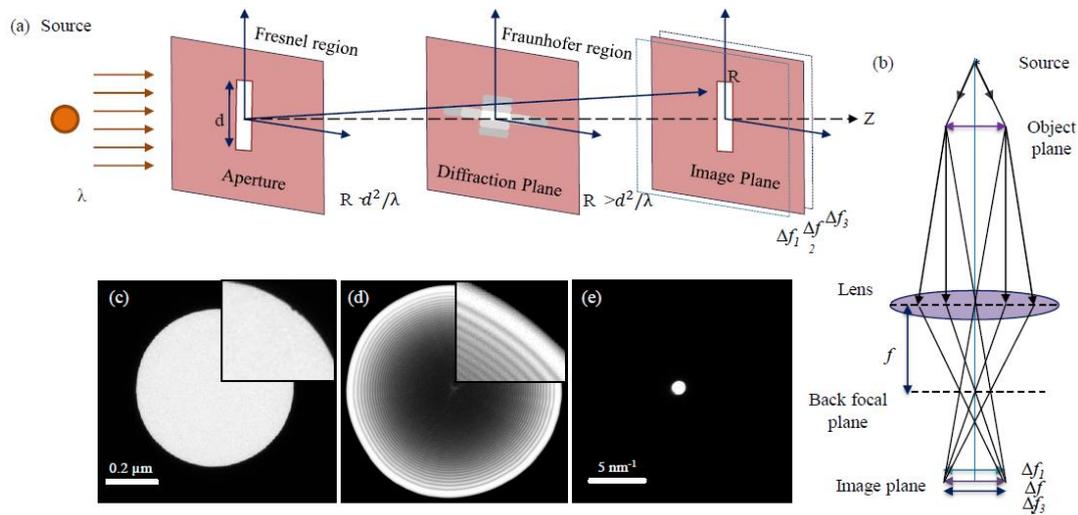

*Fig. 1.* *(a) Near field (Fresnel) and far field (Fraunhofer) regions of image formation. Also shown the exact Gaussian image plane and Fresnel regimes away from this plane. (b) Schematic showing the role of lens on the information transfer. (c)-(e) Examples of typical intensity distribution based on electron diffraction as observed at three different regions of interest, (c) at exact focus, (d) slightly away from the focus and (e) at far-field respectively. In the illustration, no specimen is used on the path of the electron beam around the optic axis.*

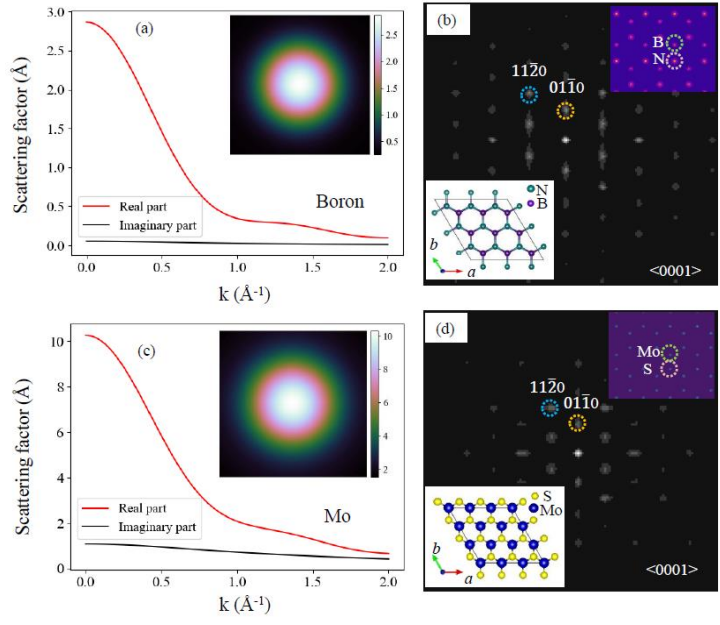

*Fig. 2.* Atom scattering factor of (a) an isolated B atom in 1D (inset 2D distribution) and (b) structure factor of monolayer BN along <0001> direction. (c) & (d) same are for Mo and $MoS_2$ lattice, respectively.

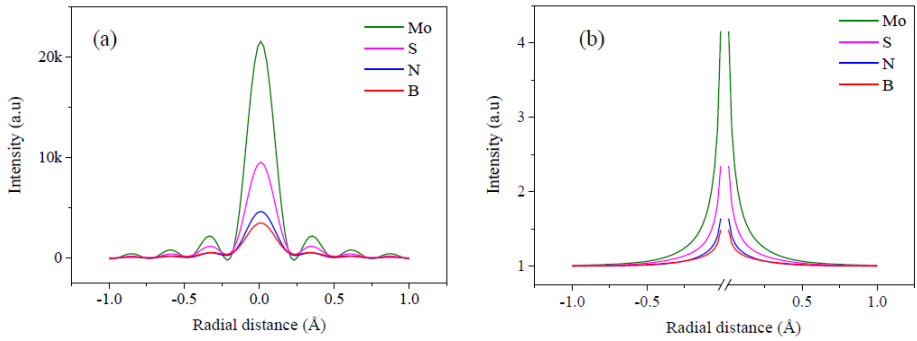

*Fig 3.* (a) image of Mo, S, N, and B atoms using Eq. 3 with Cs =-35μm, Δf =8 nm. (b) are images with Zernike like phase transfer i.e., $I(x,y) \approx 1 + 2\sigma\phi_p(-x,y)$.

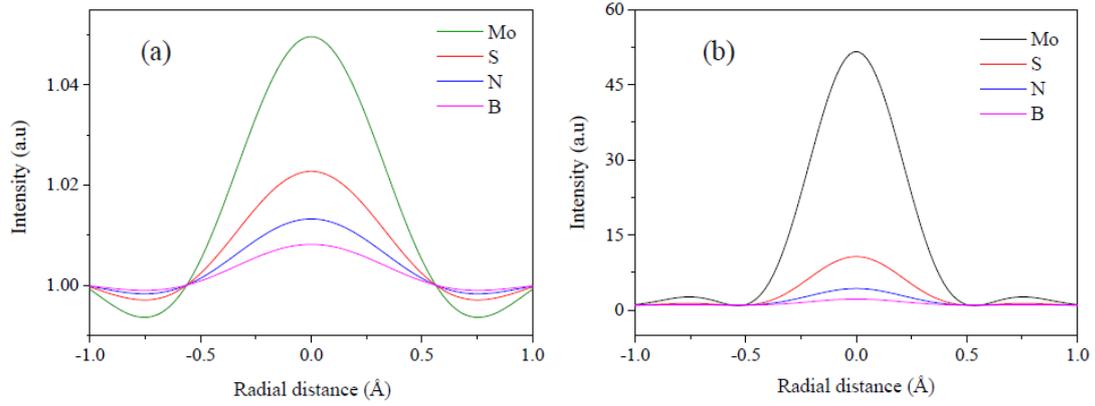

*Fig. 4.* The image of isolated (a) Mo, S, N and B calculated using Eq. 6 with $Cs = -35$ μm and $\Delta f = 8\ nm$. Only sine part of the PCTF is considered which is valid for weak phase object approximation. The effect of complete CTF on the atom image is given in (b). The outer frequency cut-off considered for the calculation is k = 0.185Å. (c).

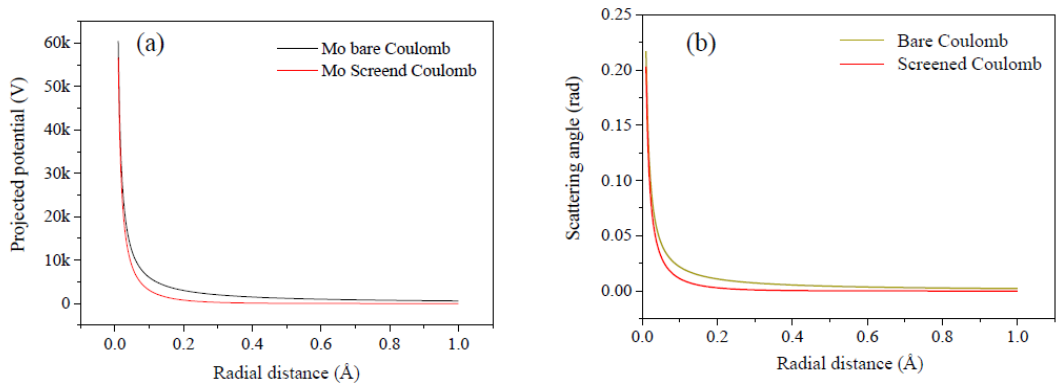

*Fig. 5.* (a) projected potential of Mo atoms, and corresponding (b) bending angle as a function of distance from the nucleus for both screened and bare Coulomb potential.

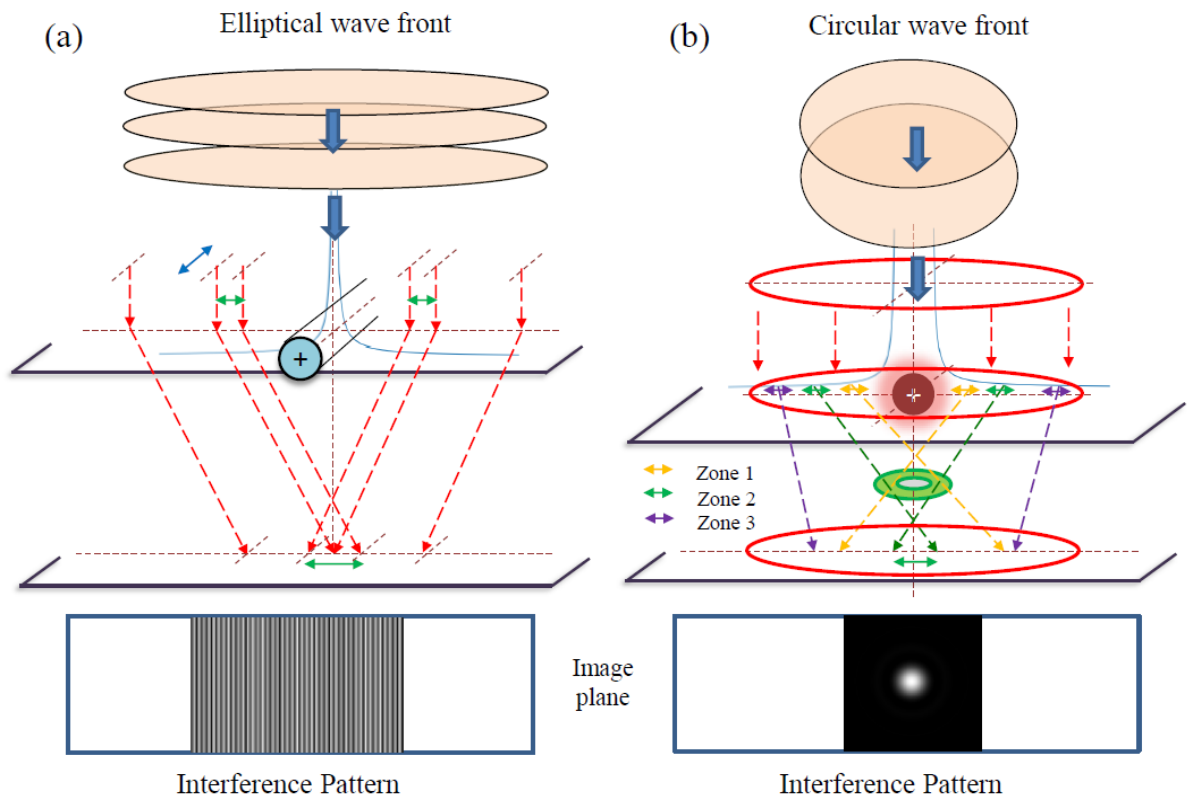

*Fig. 6.* *The interference pattern due to (a) unidirectional bi-prism and (b) atom as charge center.*

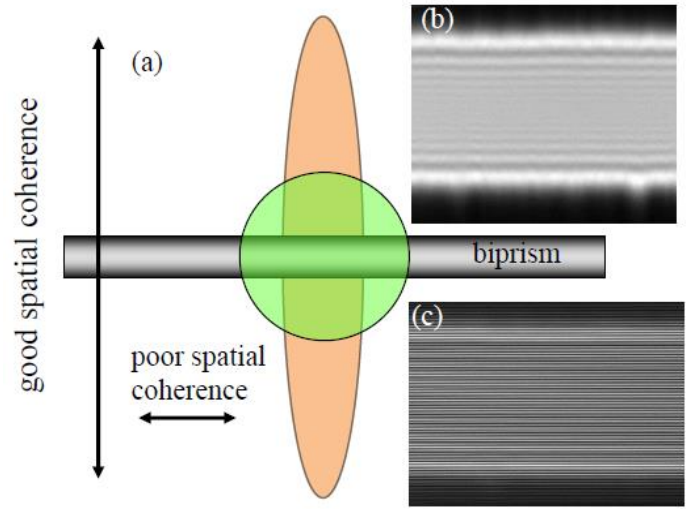

*Fig.7. (a) Schematic showing relative spread of probe for conventional round and elliptical illumination and their connection with spatial coherency. (b) & (c) fringe contrast corresponding to round and highly astigmatic (elliptical) probe in off-axis electron holography, respectively.*

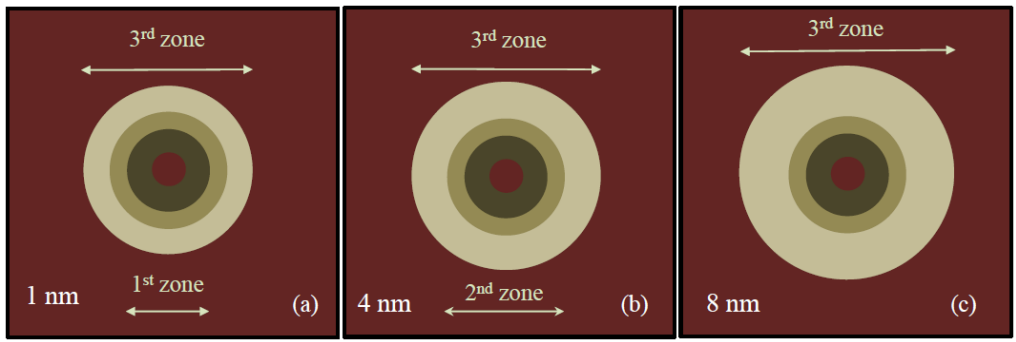

*Fig.8. (a)- (c) three different zones and their extent corresponding to 1, 4 and 8 nm focus step with respect to optic axis for Mo atom, respectively. The extent of 3rd zone increase with defocus from 10.70 pm to 21.18 pm having same mean bending angle.*

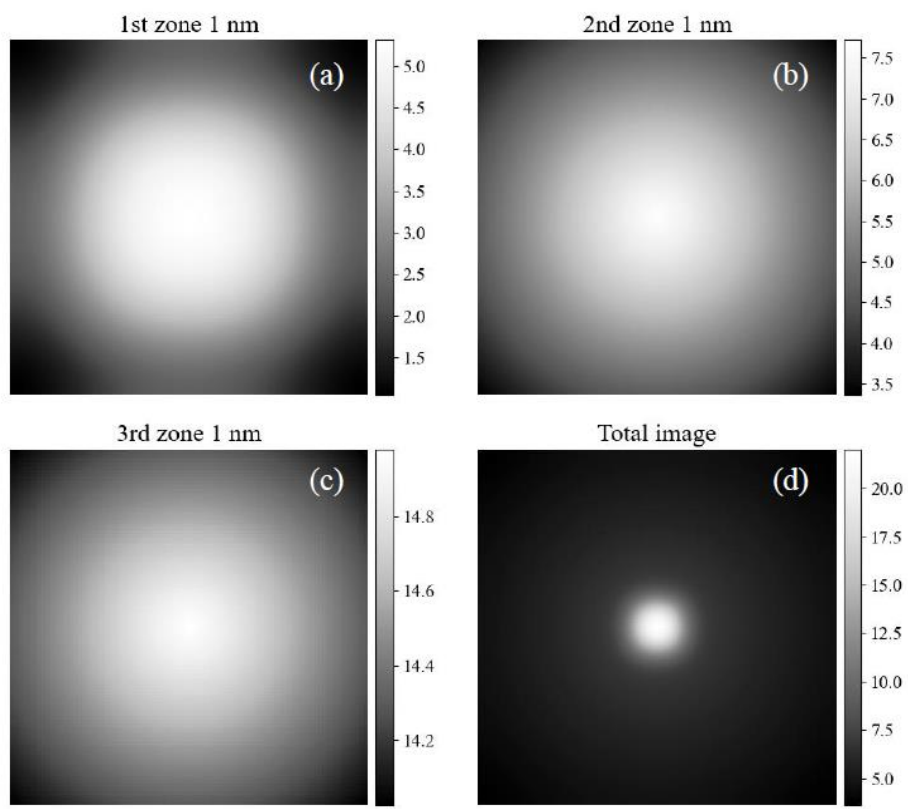

***Fig. 9.*** *(a)-(c) Image pattern formation due to various zones at 1 nm focus and (d) overall image of the Mo atom at 1 nm focus.*

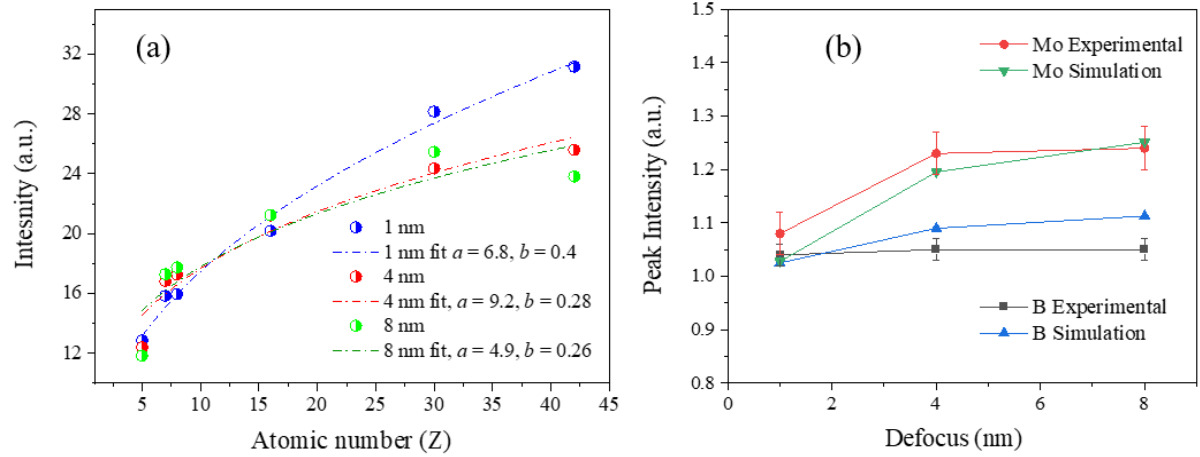

*Fig.10. Comparison of peak image contrast (a) as a function of atom number for a focus setting of 1, 4 and 8 nm without considering any aberration blurring due to $\Delta f$ and $C_s$, only diffraction blurring is effective. (b) Peak image contrast difference between Mo and B atoms after considering $C_s = -35\ \mu m$ and $\Delta f = 1, 4\ and\ 8\ nm$.*

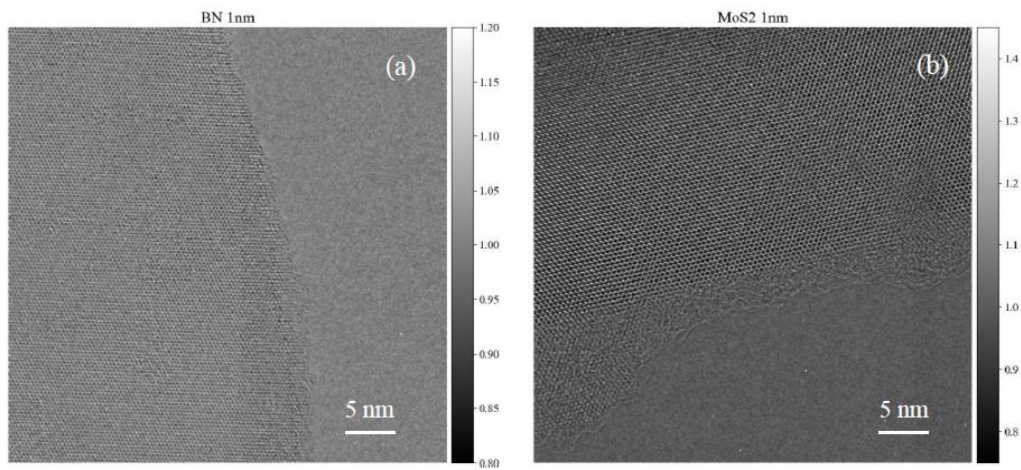

*Fig. 11. Example HRTEM image of (a) BN and (b) MoS2 at 1 nm defocus.*

# Supplementary Information

# An alternative method of image simulation in high resolution transmission electron microscopy

Usha Bhat,[1,2] and Ranjan Datta[1,2,*]

[1,2]*International Center for Materials Science, Chemistry and Physics of Materials Unit, JNCASR, Bangalore 560064.*

1. Analogy between Gabor's inline holography and HRTEM

There is a close relationship between the mathematical formalism describing Gabor's inline holography and HRTEM image formation as described within the weak phase object approximation (WPOA) and self interference of single electron wave-based description with a reference wave. Kindly see the analogy between Gabor's inline hologram, defocus HRTEM image based on WPOA and Fresnel diffraction in terms of momentum vector direction and associated interference effect in sec. S2.

A hologram is typically an interference pattern formed between incident plane wave illumination and scattered or diffracted waves due to object and is a two-step process, writing and reading [Fig. S1 (a)]. It is the intensity distribution at near field of object at some propagation distance typically in the Fresnel regime that can be captured near the Gaussian image plane through a CCD camera. In another way of stating, hologram can be considered as information contained within on how the object has diffracted the information at some propagation distance. As hologram records the complete wave field information including the effect of coherent aberration, thus it can be used to remove the effect of aberration from the image to faithfully represents the object through subsequent digital ex-situ procedures.

The amplitude of interference pattern which constitutes the ***writing component*** of the hologram is given by

$$A = \sqrt{UU^*} = \sqrt{A^{(i)2} + A^{(s)2} + 2A^{(i)}A^{(s)}\cos(\psi_s - \psi_i)} \tag{1}$$

Where, $S$ is the source, $\sigma$ is a semitransparent object, $H$ is the screen, $U = Ae^{i\psi}$ is the complex disturbance at a point in $H$, $A$ is the amplitude, $\psi$ is the phase, superscript and subscript corresponding to *(i)* and *(s)* denote incident and scattered waves, respectively.

Now, the ***reading component*** or reconstruction of the hologram is written in terms of multiplication of object transmission function with the plane wave illumination and together they form the transmitted wave function. Thus, the reconstructed wave or substituted wave is written as follows [Fig. S1 (b)]

$$U' = \alpha_p U^{(i)} = KA^{(i)2}e^{i\psi_i}[A^{(i)} + \frac{A^{(s)2}}{A^{(i)}} + A^{(s)}e^{i(\psi_s-\psi_i)} + A^{(s)}e^{-i(\psi_s-\psi_i)}] \tag{2}$$

for $\Gamma = 2$. Where $\alpha_p$ is the amplitude transmission factor of the positive, and for details on other parameters, see Ref [1].

As can be recalled that the transmitted wave function in HRTEM within weak phase object approximation (WPOA) has the form [2]

$$\psi_t(x) \sim t(x)\exp(2\pi i k_z z) \tag{3}$$

Which is similar to the ***reading component*** of the optical hologram, compare between Eq. S2 and S3. However, the description of transmission function in terms of various parameters are different.

HRTEM intensity can also be written as[3]

$$I_{in\,line} = |\psi_0 + \psi_i|^2 = A_0^2 + A_i^2 + 2A_0 A_i cos(\phi_i - \phi_0) \qquad (4)$$

Which is similar to the ***writing component*** of hologram, compare between Eq. S1 and S4.

The information about the object is carried by the transmission function component of the transmitted wavefunction. The formation of image directly from the object using appropriate lens is equivalent to form the same image from the recorded hologram with a supplementary illumination with the help of a lens [Fig. S1 (c)]. However, the mathematical expression corresponding to ***reading*** and ***writing*** *are not the same* and retaining the phase information in terms of intensity modulation is different in the context of HRTEM (see sec. 2.1, S2.6).

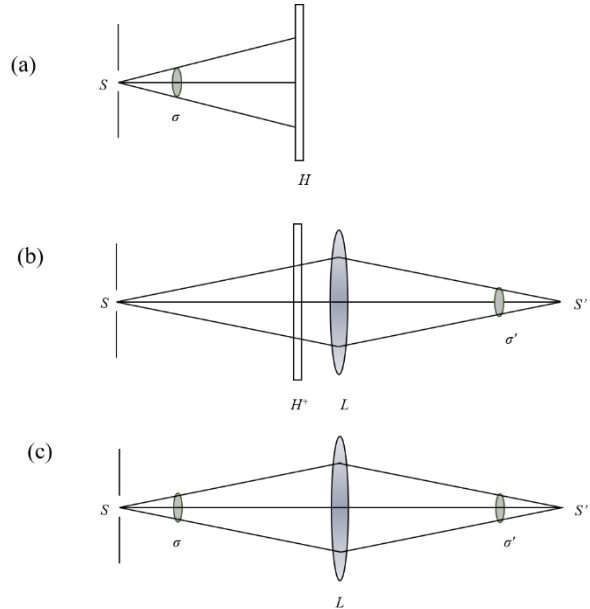

*Fig. S1. (a) & (b) are showing the writing of the hologram and object image reconstruction from the recorded hologram, respectively. (c) Equivalent image formation without a hologram. Hologram is utilized to removes the coherent lens*

*aberrations by ex-situ digital processes in order to faithfully render the image of the object.*[1]

## 2. Various methods of calculating the image intensity pattern

In the manuscript, image simulation principles based on WPOA and atom scattering (isolated atom)/structure factors (periodic atoms) are described briefly (section 2.1 and 2.2). The effect of aberration phase shift using PCTF (phase object) or ACTF (amplitude object) function is considered either in the diffraction space as frequency filter or in the image space as point spread function (*psf*). However, in diffraction space application of PCTF involves simple multiplication between the transmitted wave and aberration function and thus saves the calculation time. Otherwise, this would be rather elaborate calculation involving convolution in the real space coordinates. Fourier transformation is an essential mathematical concept to go back and forth between the intensity patterns formed in two reference planes and the results should be consistent independent of the coordinate space and methods chosen for the image calculation. We describe followings the basic procedures involved with various methods to calculate the image of atoms and different results they yield. Notes are provided to point out some of the important features associated with the formalisms and their relationship with the related methods.

### 2.1. Fresnel diffraction pattern

The Fresnel diffraction is a near field pattern perpendicular to an interface which is due to discontinuity in the scattering potential near an edge or interface.[1,4]

Huygens-Fresnel principle is based on the following two postulates

(i) Every point of a wavefront is a source of secondary disturbance giving rise to spherical wavelets and the propagation of the wavefront is regarded as the envelope of these wavelets.

(ii) The secondary wavelets mutually interfere.

Now applying the above principle which is based on the geometry as shown in Fig. S2, each surface element $dS$ of the incident wavefront $\psi_{in}$ generates a spherical wavelet contribute an amplitude $d\psi_{sc}(P)$ at a point P on the optic axis beyond the wavefront

$$d\psi_{sc}(P) = -iA(2\theta)\psi_{in}\frac{e^{ikR}}{R}dS \qquad (5)$$

after integration over all surface Eq. 5 becomes

$$\psi_{sc}(P) = -i\int_{wavefront} A(2\theta)\psi_{in}\frac{e^{ikR}}{R}dS \qquad (6)$$

After choosing appropriate limit over radial extent Eq. 6 becomes

$$\psi_{sc}(P) = i\frac{\lambda\psi_{in}^0}{r_0+R_0}e^{ik(R_0+r)} \qquad (7)$$

The Eq. S7 above describes the propagation of spherical wave. In HRTEM image simulation similar spherical wave is used in superposition with the incident plane wave front in the formulation of transmission function based on a solution of Schrödinger equation in integral form (sec. 2.2)[2,4]

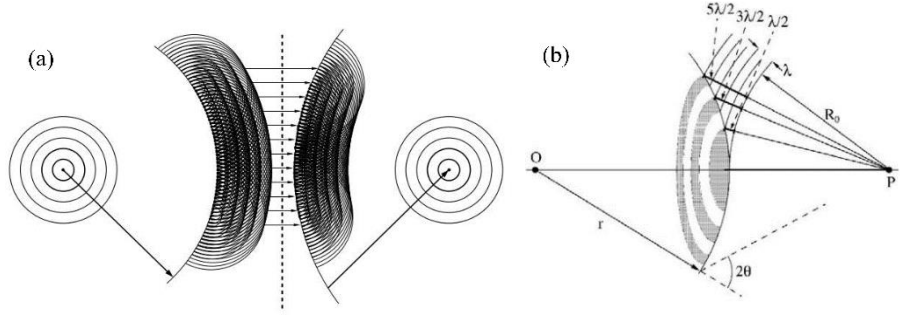

***Fig. S2.*** *(a) Geometry of Huygens principle for a diverging (left) and converging (right) wavefronts. The action of lens is along the dashed line. (b) Construction of Fresnel zones considers the self-interference of the spherical wave possessing range of momentum vectors while converging to a point P.*[4]

The Fresnel diffraction pattern for an edge can be calculated using the following equation.

$$\psi_{SC}(P) = \frac{i\psi_{in}^0 e^{ik(r_0+R_0)}}{2(r_0+R_0)} [C(X) + iS(X)]_{X_0}^{\infty} [C(Y) + iS(Y)]_{-\infty}^{\infty} \qquad (8)$$

Where *C(X)* and *S(X)* are the Fresnel cosine and sine integrals and the plot of $C(X) + iS(X)$ is called a 'Cornu spiral'.

- ❖ Therefore, one can notice that the phase change in the propagating spherical wave is due to range of angular momentum vector directions (same path length lies on the surface of the sphere) associated with a spherical wave (for plane wave it is only one momentum direction) and the self-interference while converging between those various angular momentum vector directions. Various angular momentum vector directions will acquire path difference between themselves due to *outside curvature of spherical surface* with respect to the converging point at ***P*** and the problem of interference is solved by well-known Fresnel zone construction. More precisely, the wave fronts geometries are spherical, parabolic, and plane surface for Rayleigh-Sommerfeld, Fresnel and Fraunhofer regimes, respectively. The rate of change of phase between wavevectors that is governed by the outward curvature of the wave front, is different for three different regimes. A similar picture is also captured in self-interference between probe and scattered waved that describes the HRTEM image intensity pattern (see Eq. S4 and Ref. [4,6]).[5,6] The momentum vector direction of $\psi_0$ is along the $k_0$ direction, and $\psi_i$ along various $k$ directions. Kindly note that the interference geometry is different for Fraunhofer pattern and off-axis electron holography (see S2.2 & S2.5).

**2.2. Fraunhofer diffraction pattern**

Fraunhofer integral can be employed for various geometry of slits to calculate intensity pattern in the diffraction plane as a function of scattering angle based on analytical expressions.

As an example, the Fraunhofer integral for a rectangular aperture of sides 2a and 2b with origin at the center O of the rectangle and with $O\xi$ and $O\eta$ axis parallel to the sides is given as follows

$$U(P) = C \int_{-a}^{a} \int_{-b}^{b} e^{-ik(p\xi+q\eta)} d\xi d\eta = C \int_{-a}^{a} e^{-ikp\xi} d\xi \int_{-b}^{b} e^{-ikq\eta} d\eta \qquad (9)$$

The intensity is given by

$$I(P) = |U(P)|^2 = \left(\frac{\sin kpa}{kpa}\right)^2 \left(\frac{\sin kpb}{kpb}\right)^2 I_0 \qquad (10)$$

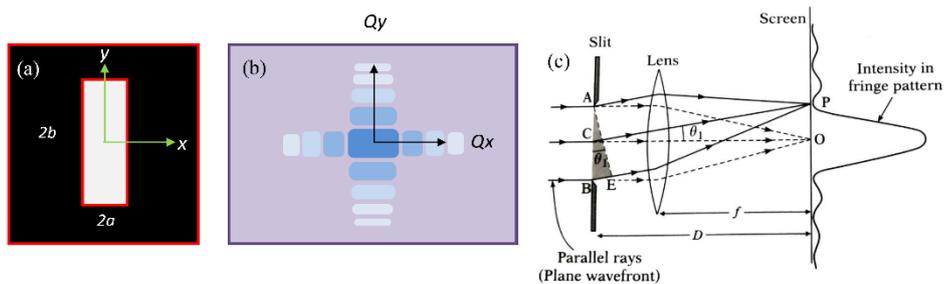

***Fig. S3.*** *(a) Geometry of rectangular aperture and (b) corresponding far field or Fraunhofer intensity pattern. (c) Geometry of Fraunhofer interference showing the origin of path difference.*

Experimental observation of example Fresnel pattern and far field Fraunhofer pattern are given in Fig S4.

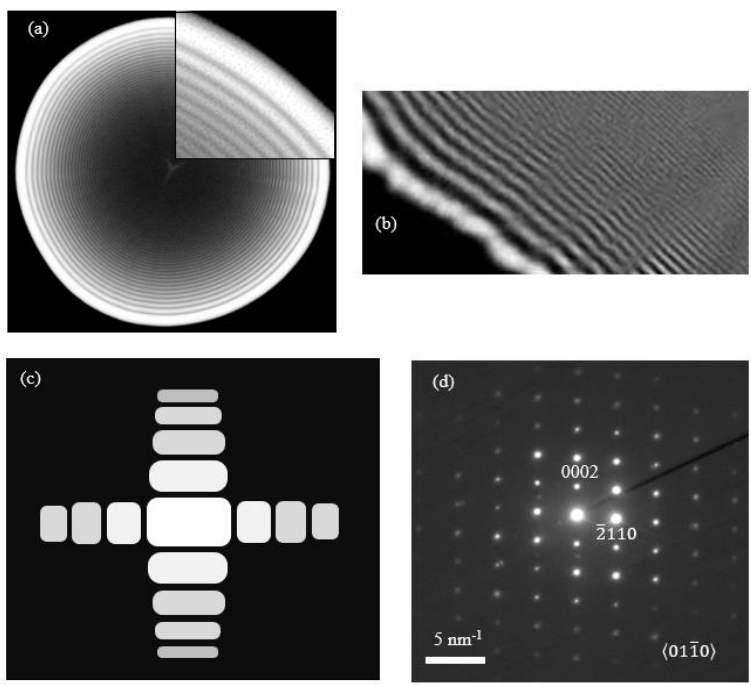

***Fig. S4.*** *Fresnel diffraction pattern as observed under slight defocus (~ 10s of nm) condition from the (a) edge of an aperture, and (b) thin specimen edge. (c) Far field*

*Fraunhofer pattern from single slit and (d) electron diffraction pattern of ZnO Crystal along <01-10> Z.A.*

- ❖ As already mentioned that the interference geometry for the far field pattern is different than the Fresnel zone construction. In this case the correlation of path difference between emerging waves from various spatial points at the aperture plane is considered. The path difference is not between different momentum vector directions rather spatial separation between two points having same momentum or scattering directions. In case of off-axis electron holography the interfering waves have momentum direction mirror symmetry to each other (sec. S2.5).

**2.3. Fourier transformation-based method for calculating image pattern**

In the preceding two sections, the propagation of phase information from the aperture plane to the near field regime (Fresnel) diffraction plane (Fraunhofer) are briefly described along with emphasis on specific diffraction geometry of waves depending on the momentum vectors. The change in momentum vector directions of the scattered/emergent waves and concomitant interference phenomena are at the heart of calculating the intensity pattern.

On the other hand, the Fourier transformation (FT)-method is based on the principle that the intensity pattern corresponding to image and diffraction planes are related by FT without considering above mentioned diffraction geometry. For example, diffraction pattern of a periodic crystal oriented along high symmetry direction is the Fourier transformation of the periodic crystal potential. However, the independent variable they represent are different, in case of FT the independent variable is frequency of wave vector $k$ and for Fraunhofer analytical formula it is scattering angle $\theta$. The abs-FT pattern needs to be calibrated with respect to $\theta$ or reciprocal lattice vector $g$. The mathematic operation involved in Fourier method has the similarity with the physical picture of Abbe's theory of imaging [sec S2.4.] for more details on Abbe imaging theory and its connection with off-axis electron holography formalism in sec. S2.5.

Now, according to the Fourier method, the far field Fraunhofer pattern is obtained by Fourier transformation of the object function $f(x)$ (or $f(x, y)$ in 2D) as

$$\mathcal{F}(k) = \int_{-\infty}^{\infty} f(x) e^{-2\pi i x \cdot k} dx \qquad (11)$$

Strictly speaking, it is the magnitude of $\mathcal{F}(k)$ or absolute FT, which results in the Fraunhofer pattern at far field and not the individual $\boldsymbol{Re(x,k)}$ and $\boldsymbol{Im(x,k)}$ components.

$$\boldsymbol{Abs\ (or\ modulus\ or\ magnitude)\ \mathcal{F}(k) = Fraunhofer\ pattern} \qquad (12)$$

However, it is indispensable to know the $\boldsymbol{Re(x,k)}$ and $\boldsymbol{Im(x,k)}$ components in order to go back to the original function $f(x)$ through inverse Fourier transformation. The established phase retrieval procedures involve retrieving the object wave function (OEW) through appropriate filter function applied directly on $I(k,z)$ derived from the image intensity $I(x,y,z)$ recorded at sufficient resolution.[6–8] Experimentally, if only magnitude of $\mathcal{F}(k)$ is recorded in the diffraction plane, then $\boldsymbol{Re(x,k)}$ and $\boldsymbol{Im(x,k)}$ components are lost, and this is the well-known phase problem. Though it is not a problem if one derives $\mathcal{F}(k)$ or $I(k,z)$ from the recorded intensity pattern in the image plane with sufficient spatial resolution and the phase related to object potential and crystallography are preserved and can be retrieved by following certain procedures.

To elaborate more, what happens in the FT process is that the object function at first is expanded into its continuous cosine and sine series with frequency ranging from $k = -n\ to\ +n$ as written below

$$Re(k) = \int_{-\infty}^{\infty} f(x)\cos(2\pi x.k)dx \qquad (13)$$

and

$$Im(k) = \int_{-\infty}^{\infty} f(x)\sin(2\pi x.k)dx \qquad (14)$$

$$Abs\ square\ of\ \mathcal{F}(k) = \left|\sqrt{Re^2 + Im^2}\right|^2 \qquad (15)$$

$$and\ the\ phase\ \theta(k) = \tan^{-1}\frac{Im}{Re} \qquad (16)$$

The expansion of the object function (discrete or periodic) into various cosine and sine harmonics of different frequencies resulting in Fourier waves with various frequencies and are similar to Abbe waves [sec. S2.4]. The absolute FT is then plotted for each frequency and the result is the well-known Fraunhofer pattern. Moreover, each frequency of Fourier wave is associated with corresponding phase. This phase as a function of $k$ can be calculated [Fig.S5 (d)].

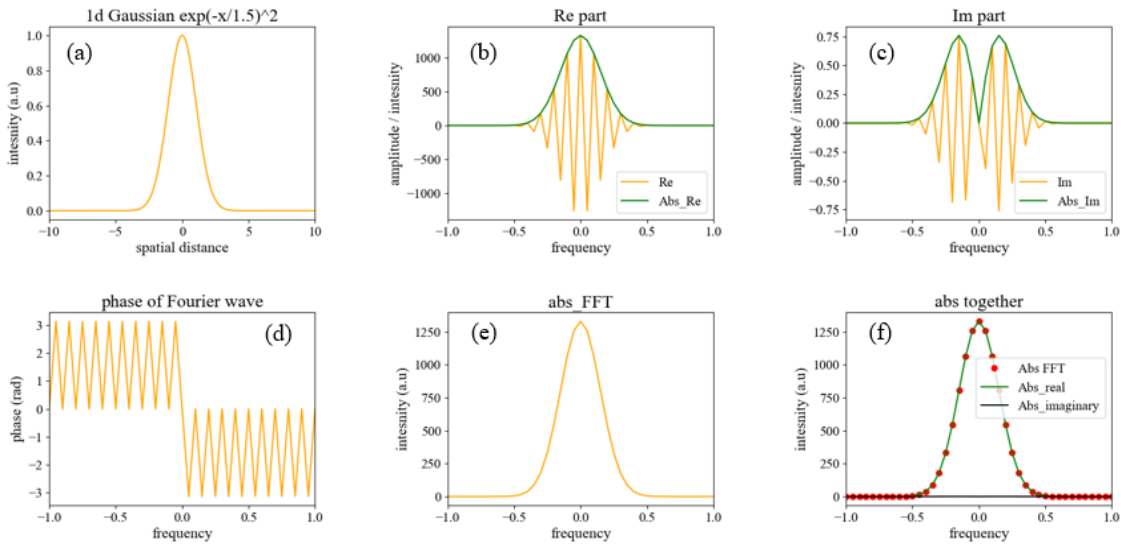

***Fig. S5.*** *(a) example of a Gaussian function $y = exp(-\frac{x}{1.5})^2$, (b) & (c) corresponding **Re** and **Im** parts along with respective absolutes after Fourier transformation, respectively. (d) Phase of Fourier waves as a function frequency, (e) Absolute of FT, and (f) plotting together total abs-FT, abs-Re and abs-Im, showing that abs-Im is negligible and abs-Re is almost equal to abs-FT.*

The implication is that for a discrete function, e.g., if the function is a narrow slit, it will require higher magnitude of frequency to get a first zero in the frequency pattern compared to wider slit. The subsidiary effect is due to non-perfect cancellation of integrand depending on frequency. For periodic function or slit, similar explanation holds. Now, as already mentioned that in this Fourier based method one adjustment is required that is to calibrate or relate various frequencies with the scattering angle, which do not arise naturally. This can only be done i.e., to relate the frequencies at which magnitudes of Abs FFT are zeroes with the angle at which destructive interference takes place between waves at two spatial points in the slit/object level. The later aspect is the physical picture required to derive the analytical formula based on Fraunhofer integral [sec S2.2.]. The results calculated independently by FT and analytical approach are shown in Fig. S6 for example seven periodic slits and have little difference between them.

The analytical formula for various slit geometry is derived based on path difference of waves between two spatial points at the slit opening emanating along various direction based on the principle of constructive and destructive interference. The method has the advantage over Fourier based method in the sense that the formula contains useful quantities e.g., wavelength (λ) of the illumination, scattering angle (θ), dimension and periodicity of the slit.

For diffraction from N number of periodic slits the intensity expression is given by

$$I(\theta) = I_0 \left[\frac{\sin(\frac{\pi a}{\lambda}\sin\theta)}{\frac{\pi a}{\lambda}\sin\theta}\right]^2 \left[\frac{\sin(\frac{N\pi d}{\lambda}\sin\theta)}{\sin(\frac{\pi d}{\lambda}\sin\theta)}\right]^2 \tag{17}$$

Where, $a$ is the slit width, $d$ is the inter-slit distance or periodicity. $\lambda$ is the wavelength of the monochromatic wave.

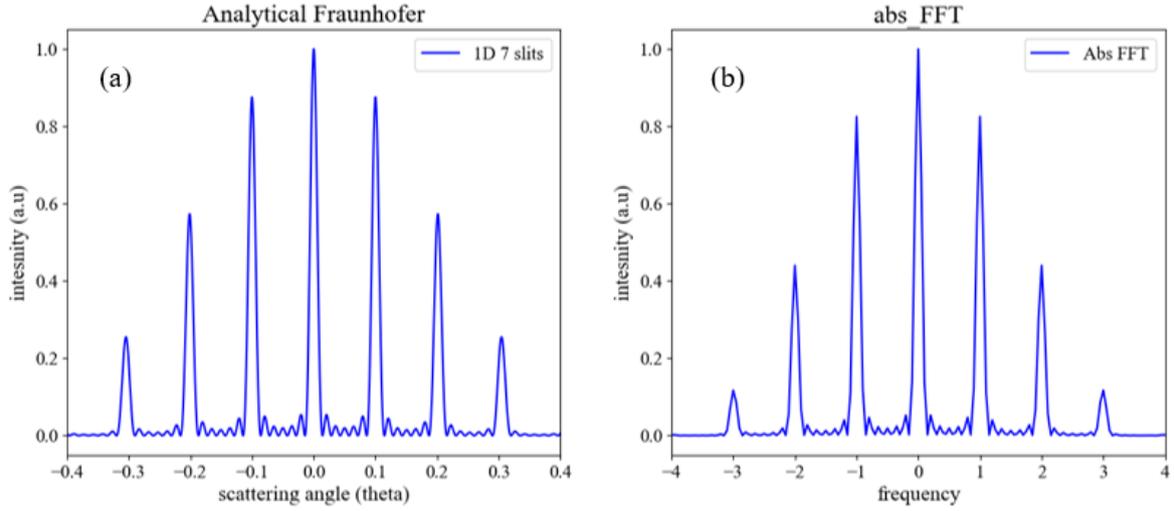

*Fig. S6. (a) comparison of Fraunhofer pattern calculated by Fraunhofer analytical methods for periodic 7 slits with slit size a = 2λ and periodicity d =10 λ and (b) Fourier transformation of 7 periodic slits followed by abs-FT. (b) appearance of zero magnitude in abs FFT is due to zero values of integration of $Re(x,k)$ and $Im(x,k)$ parts with respect to x corresponding to respective frequency values.*

❖ **Note on $k$:**

This physically represents that each object point will give off **Re** and **Im** parts of the Fourier wave with frequencies $k = -n\ to\ +n$. For each frequency one need to evaluate the integrand corresponding to **Re** and **Im** part and sum up to obtain the abs-FFT for that frequency. Therefore, in the process of evaluating the **Re** and **Im** part, the amplitude and phase of each Fourier waves i.e., the numbers outside and inside the cosine and sine part mixed up. Or in other words they collapse to a single number. Moreover, the amplitude of each cosine and sine part will depend on the values of corresponding cosine and sine multiplied with $f(x)$. This is how mathematical procedure mix up the information on function and frequency of waves. In HRTEM and off-axis holography image reconstruction, the starting data is image, therefore, every step while performing FT is preserved and then only it is possible to go back to the image back again.

**2.4. Note on Abbe's theory**

According to Abbe's theory, it is postulated that the point of interaction between illuminating monochromatic wave and object gives rise to emerging waves with continuous spectrum of frequencies. Higher the frequency of wave higher is the scattering angle. The waves of equal frequency and converging angle meet (by ideal microscope lens) to produce standing waves corresponding to different discrete frequencies. These standing waves superpose to form image wave function. Largest frequency component which is allowed by the aperture defines the resolution according to Abbe's famous formula of resolution. These standing waves are exactly similar to the Fourier transformed waves discussed above except object function being considered and form the basis of Fourier optics. The diffraction geometry is given in Fig. S7 and showing interference between pair of wave vectors having same inclination angle. There is a continuous presence of such pairs of partial waves with continuous range of inclination angles. This is similar to off-axis electron holography only for a pair of momentum vectors across a mirror plane.

Now, the resolution criterion is derived as follows. The objective lens acts as low-pass filter. The cut-off frequency in 1D is given by

$$k_{max} = \frac{2\pi}{\lambda f_{obj}} x_{max} = \frac{2\pi}{\lambda f_{obj}} \theta_{max} \qquad (18)$$

The effect of limiting frequencies higher than maximum allowed by the aperture is to limit the resolution of the system.

The impulse response or point spread function (*psf*) due to limiting objective aperture is given by the Fourier transformation of objective aperture to the image plane in 2D

$$g(r) = \frac{J_1(k_{max} r)}{k_{max} r} \qquad (19)$$

Where, $r = x^2 + y^2$

Now the Rayleigh's criteria for resolution is postulated in terms of minimum distance between two such blurred points due to aperture. This is defined by the first root of two such function i.e. the maximum of one function overlaps with the root of second function. This occurs at $x_o = 1.22\pi$

The resolution is given by

$$\rho_0 = 0.61 \frac{\lambda}{NA} \qquad (20)$$

Where, $NA = \frac{r_{max}}{f_{obj}}$ is the numerical aperture, represents the maximum half-angle subtended by the entrance pupil.

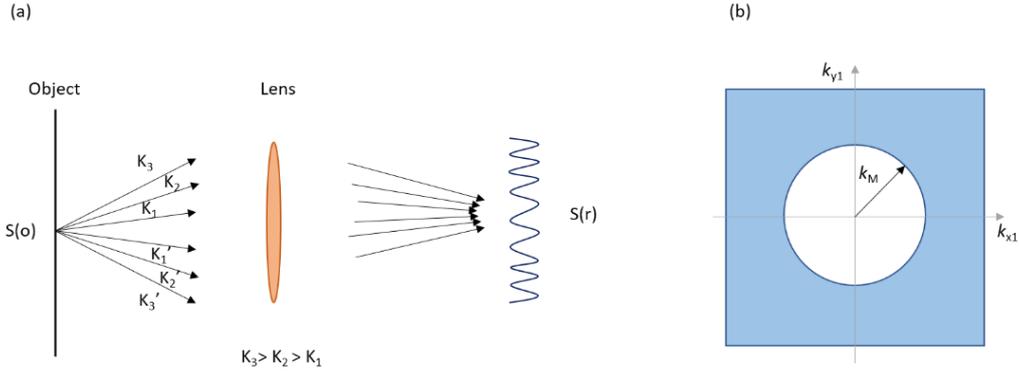

*Fig.S7. (a) Abbe's picture of image formation showing range of frequencies generated at the point of interaction between illumination wave and object point, (b) Maximum frequency allowed by aperture will define the resolution of the image.*

### 2.5. Image intensity pattern in off-axis electron holography

Now there is another way to calculated pattern or image based on converging waves. This is based on geometry of interference associated with off-axis optical/electron holography. The equation for each frequency in 2D is given by (Lehmann 2008) [9,10]

$$I_{hol}(\vec{r}) = I_0 + I_{ima}(\vec{r}) + 2|\mu|A_0 A(\vec{r})\cos(2\pi \vec{q_c}\cdot\vec{r} + \phi(\vec{r})) \qquad (21)$$

Where, $I_0 = A_0^2$ is the intensity of reference wave. $I_{ima}(\vec{r})$ is the image intensity and $q_c = k\beta$ is the spatial carrier frequency of the hologram. $\mu$ is the degree of coherence.

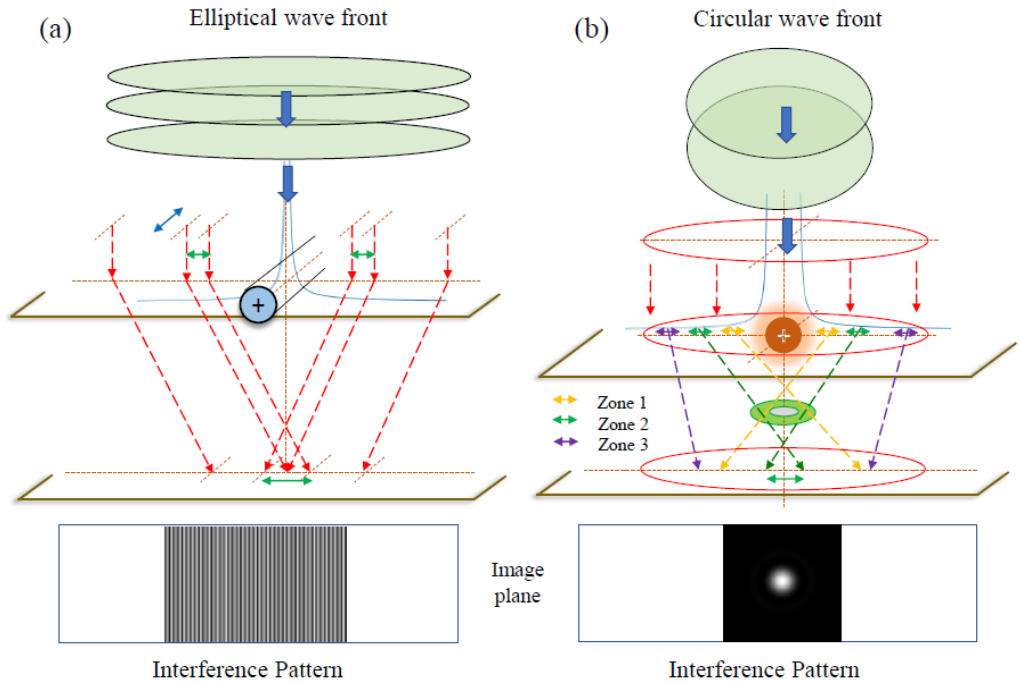

***Fig.S8.*** *(a) Image formation in off-axis electron holography with wave of single frequency having momentum vector direction equal and mirror symmetry with respect to biprism axis. (b) electron interference along radial direction.*

Now there is a connection between Abbe's model and the carrier frequency associated with the off-axis electron holography. According to Abbe, waves intercepted by aperture at higher scattering angle is associated with higher frequency, whereas higher angle of interference in case of off-axis geometry is associated with higher carrier frequency. Both gives finer details and associated with better spatial resolution.

In fact, the new simulation approach proposed in the present manuscript is based on the off-axis electron holography concept by considering single atom as an electrostatic interferometer [Fig. S8]. Kindly see sec. S2.9 and sec. 2.3 of manuscript for more details. The geometry of interference is different in this case compared to Fresnel and Fraunhofer cases (see Fig. S8, S2 and S3 for comparison). In case of off-axis electron holography it is the pair of momentum vectors across the mirror plane which produces interference pattern.

**2.6. Approach based on weak phase object approximation (WPOA)**

This is the widely applied approach to calculate phase contrast images in transmission electron microscopy along with the consideration of imperfect lens acting as phase plate. The transmitted wave function within kinematical scattering is given by

$$\psi_t(x) \sim \exp(2\pi i k_z z) \exp(i\sigma V_t z) \qquad (22)$$

Where, $V_t$ is the projected specimen potential and σ is the interaction parameter and not the scattering cross section, and is given as

$$\sigma = \frac{2\pi}{\lambda V}\left(\frac{m_0 c^2 + eV}{2m_0 c^2 + eV}\right) = \frac{2\pi m e \lambda}{h^2} \qquad (23)$$

The transmitted wave function can now be written as

$$\psi_t(x) \sim t(x) \exp(2\pi i k_z z) \qquad (24)$$

The transmission function is given by

$$t(x) = \exp[i\sigma V_t(x)] \qquad (25)$$

Now invoking WPOA, i.e. ignoring the terms with $\sigma^2$ and higher order, transmission function is approximated to

$$t(x) = 1 - i\sigma V_t(x, y) \qquad (26)$$

This is an important approximation which is equivalent to Zernike phase object [sec. S2.6.2]. Moreover, there is a close resemblance between Gabor's reading component of in-line hologram and the expression in Eq. S23. This similarity is the basis for terming HRTEM images as in-line holograms. [sec. S1]. However, the above description of WPOA does not draw any analogy between Gabor's in-line holography, Fresnel diffraction geometry and defocus HRTEM image in terms of interference geometry. However, instead considering the change in momentum vector direction due to interaction with the object potential draws analogies and differences all the three pictures in terms of geometry of interference (see notes in supp. S 1&2).

Now considering lens response, the image wave function becomes

$$\psi_i(x,y) = 1 - i\sigma\phi_p(-x,y) * \mathcal{F}\{P(u,v)\exp(i\chi(u,v))\} \tag{27}$$

And the image intensity

$$I(x,y) = \psi_i(x,y)\psi_i^*(x,y) \approx 1 + 2\sigma\phi_p(-x,y) * \mathcal{F}\{sin\chi(u,v)P(u,v)\} \tag{28}$$

Where,

$$\sigma\,\mathcal{F}\{sin\chi(u,v)P(u,v)\} = \frac{2\pi}{\lambda^2}\int_0^{\theta_{ap}} sin\chi(\theta)J_0(\frac{2\pi\theta r}{\lambda})\theta d\theta \tag{29}$$

The above intensity expression (Ref.[11] -pp 62 ) only in terms of sine convolution is obtained by omitting $\sigma^2$ term (Ref. [12] -pp 487 ).

Now, Eq. S28 describes the point spread function (*psf*) in the image plane. Similar expression derived by Scherzer,[13] Kirkland,[2] incorporates the atom scattering factor and the transmitted wave function was derived based on Fraunhofer approximation and Schrödinger integral equation, respectively.

Now to see how the *psf* acts on the image contrast, one needs to take a closer look in Eq. S28, Eq.6, S37 and S18. The Bessel function term has the origin in aperture function that sets the resolution criteria according to Abbe's theory. This term will control the image pattern width in terms of FWHM. The sine function, atom scattering factor and interaction constant $\sigma$ will contribute to the weight to the image intensity, as magnitude of phase contrast, atomic number and characteristics of probe electron, respectively. The effect of atomic number enters into Eq. S26 (based on WPOA) as atomic potential. Though all these expressions are equivalent but give different image pattern depending on how the transfer function is considered during calculation (see Fig. 3 & 4 in main manuscript for comparative results). Kindly note that the aberration phase shift is never added as additional phase inside the trigonometric operator of wave functions either in image or diffraction plane, rather it acts as frequency filter in the diffraction plane and *psf* through convolution in image plane. If this phase is added, then it will cause change in the pattern periodicity and is never observed. See the primary text for comparison of images for the same atoms by various formalism.

### 2.6.1. Note on the series approximation in Eq. S25

In the transmitted wavefunction, the information about the object in terms of projected potential $V_t(x, y)$ is carried by the transmission function in Eq. S25. The quantities inside the exponential of the transmission function is the change in phase and can be added with the plane wave phase of the illumination as follows or kept inside the exponential operator as it is if we ignore $\exp(2\pi i k_z z)$ term from the expression as given in Eq. S23 ~~as was done in many occasion~~.[6,14]

$$\psi_t(x) \sim \exp\bigl(i(2\pi k_z z + \sigma v_z(x))\bigr) \quad (30)$$

However, if we don't do any such series approximation then, phase information will be inside the cosine and sine trigonometric functions and intensity of transmitted radiation derived by multiplying the transmitted wave function with its complex conjugate will result in a constant value and thus phase information is lost. Therefore, by expanding the transmission function in a series (weak phase object approximation) the phase is recovered and multiplying with complex conjugate will retain the phase. Thus, we can say that by mere mathematical manipulation information on object phase is retained.

So, if we don't do the series expansion and approximation then

The image wave function can be written as

$$\psi_i(x,y) = \{t(x)\exp(2\pi i k_z z)\} * \mathcal{F}\{P(u,v)\exp(i\chi(u,v))\} \qquad (31)$$

Image intensity

$$I(x,y) = 1 \qquad (32)$$

See derivation below, let

$$p = 2\pi k_z z + \sigma v_z(x) \qquad (33)$$

$$\psi_i(x,y) = \{\cos(p) + i\sin(p)\} * \{\cos(x) + i\sin(x)\} \qquad (34)$$

$$I(x,y) = \psi_i(x,y)\psi_i^*(x,y) \approx 1 \qquad (35)$$

However, there is a way out of the series approximation or WPOA. To read this additional phase shift we need to have a reference wave with respect to that the fringe shift will be visible and change in phase can be measured as was done in Eq. S4 and Eq. 12 in Ref. [6]. This is based on the physical picture of self-interference and the intensity expression looks similar to expression corresponding to off-axis electron holography except the additional carrier frequency and writing component of Gabor's holography.[5]

### 2.6.2. Note on transmission function and phase object:

The definition of a transmission function can be found in (Ref. [15] -pp 446-447 ). Any diffraction grating which introduces variation in amplitude and phase on the incident wave can be characterized by its transmission function. It is given by

$$F(\xi, \eta) = \frac{V(\xi,\eta)}{V_0(\xi,\eta)} \qquad (36)$$

Where, $V_0(\xi, \eta)$ is the disturbance on $(\xi, \eta)$ plane in the absence of object and $V(\xi, \eta)$ is the disturbance on the same plane when object is present. This is consistent with the formalism of transmission function for electron imaging discussed before.

In page 472 of Born and Wolf, in Zernike's phase contrast method section, the 'phase object' is defined by a complex amplitude function (for light) as follows

$$F(x) = e^{i\phi(x)} \qquad (37)$$

Where, $\phi(x)$ is a real periodic function and whose period is equal to the period of grating [in case of periodic grating]. For $\phi(x)$ small compared to unity the above equation can be approximated to

$$F(x) \sim 1 + i\phi(x) \qquad (38)$$

Now one can notice the origin of observing $\phi(x)$ in the intensity information. If this approximation is not done, $\phi(x)$ would be lost. This is similar to the discussion already made in the context of WPOA and transmission function.

### 2.6.3. Note on convolution

The convolution operation of the transmitted wave function with point spread function results in very unusual behavior of atom image with slight variation in lens parameter and aperture angle. See some of the figures below;

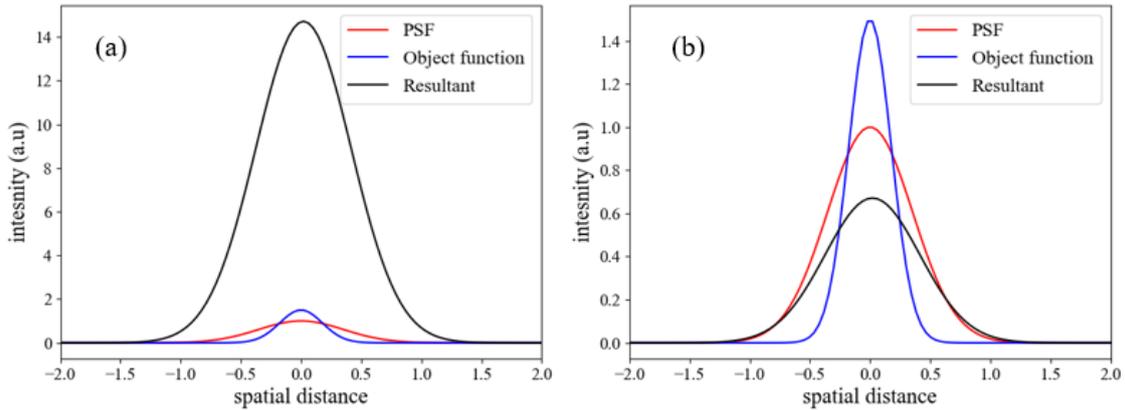

*Fig. S9. (a) Examples showing the area under the graph (resultant black) is not preserved after convolution between a model object function (Gaussian blue, with peak value 1.5 and FWHM 0.5) and convoluting function (Gaussian red, with peak value 1 and FWHM 1). (b) Resultant black is normalized based on the total area under the curve of model object function (flux balance) which then shows the broadening at the expense of reduced peak intensity.*

This convolution operation changes the peak height of the resultant curve significantly. Therefore, one need to normalize the resultant based on flux balance i.e. area under the object function. After considering flux balance the resultant curve drops in peak value and delocalizes (blurring) compared to the object function. Experimentally acquired images need to be deconvoluted with the known *psf* and compare the image with the simulated image.

However, any flux balance approach will be relative between atoms and for absolute contrast the present alternative method may be used.

**2.7. Note on image intensity calculation based on Moliere scattering factor**

The Eq. 6 in the main manuscript has some precursor work. [13,16] For example, in Scherzer's theory of phase contrast, phase shift is introduced by CTF function in the back focal plane and the wavefunction of the object in image plane at Fraunhofer approximation is given by

$$\psi_i(r_i, \nu_i) = \frac{1}{\lambda_i f_i M} \int_0^{r_m} r dr \int_0^{2\pi} d\nu \times \exp\left(2\pi i \frac{r_i \lambda \cos(\nu-\nu_i)}{\lambda_i f_i M}\right) \psi_a(r, \nu) \tag{39}$$

$$\psi_i(r_i) = \frac{2\pi}{M\lambda} \int_0^{\alpha} \exp[i(s^2\Theta^4 - \tau s\Theta^2] \times S(\Theta) J_0\left(\frac{2\pi r_i}{M\lambda}\Theta\right) \Theta d\Theta \tag{40}$$

In the publication by Eisenhandler and Siegel[16] similar to the above formalism, the image wave function and corresponding intensity are given by

$$\psi(x^i, y^i) = \frac{2}{M\lambda} \int_0^{\pi} \int_0^{\alpha_{max}} \psi(\alpha, \theta) e^{i\chi(\alpha,\theta)} \times \cos\left(\frac{2\pi x^i \cos\theta}{\lambda}\right) \cos\left(\frac{2\pi y^i \sin\theta}{\lambda}\right) \alpha d\alpha d\theta \tag{41}$$

$$|\psi_{total}(x^i, y^i)|^2 \approx M^{-2} + 2M^{-1} Re[\psi(x^i, y^i)] \tag{42}$$

In the above formulations of equations (S40) and (S41) following considerations are made; Some of the illuminating monochromatic electron wave after passing through the sample scatter elastically. The scattered wave undergoes phase change and amplitude of scattered wave is a function of scattering angle. The intensity distribution at the diffraction plane is the Fraunhofer pattern which is obtained by Fourier transformation of object plane distribution. The amplitude distribution and phase relationship are important to go back to the image plane by inverse Fourier transformation. Now it is considered that each point of diffraction pattern emits Huygens wavelets and the wavelets which can escape the aperture at the back focal plane will interfere/recombine at the image plane to form the image. This can be obtained by inverse Fourier transformation and any information lost is due to wavelets propagated outside the aperture. Equation (S40) and (S41) are the final image wave function and intensity calculated accordingly for an aberrated lens.

The above formalisms are equivalent except some constant factor like $r_i/\lambda$ in the Scherzer equation. For periodic lattice atom scattering factor will be replaced by corresponding structure factor.

## 2.8. Review notes on fundamental concepts of wave, associated phase, and phase shift

The change in phase is the fundamental aspect behind interference and formation of contrast in the image. However, this has been treated differently by different methods. Therefore, in the following, discussion is provided, considering only one-dimensional case to some useful concepts on parameters related to plane wave function which are used to calculate useful quantities i.e., phase and amplitude of object exit wave function (OEW) and intensity pattern of HRTEM image.

A wave is parameterized by its amplitude ($A$), phase ($\varphi$) and wave vector ($k = 1/\lambda$). It takes the form of a real sine or cosine functions mimicking the profile for example of a water wave, periodic sound signal, sound wave or vibrations etc. in a physical medium.

$$U(x,t) = A(x,t)\cos(kx - \omega t \pm \varphi) \qquad (43)$$

The sine form is merely the phase shifted cosine form by π/2.

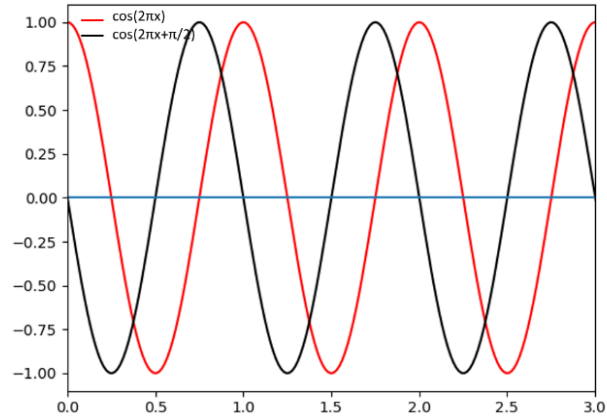

*Fig. S10. Schematic of a standing wave based on real cosine function. Sine function is the π/2 phase shifted cosine function.*

However, plane wave of light and electron is expressed by the complex wave function as given below (omitting the time variable *t*) that is consistent with the quantum mechanical type wave function.

$$\psi(x) = A(x)\, exp\, i(2\pi kx \pm \Delta\varphi) = Re + Im \qquad (44)$$

Where, $Re = A(x) \cos(2\pi kx \pm \Delta\varphi)$ and $Im = A(x) \sin(2\pi kx \pm \Delta\varphi)$

The relationship between phase difference with path difference is

$$\Delta\varphi = \frac{2\pi}{\lambda} \Delta x \qquad (45)$$

Thus equation (S7) can also be written as for a complex wave

$$\psi(x) = A(x)\, exp\, i[2\pi k(x \pm \Delta x)] = Re + Im \qquad (46)$$

The *Δφ* can either be a scaler or vector.

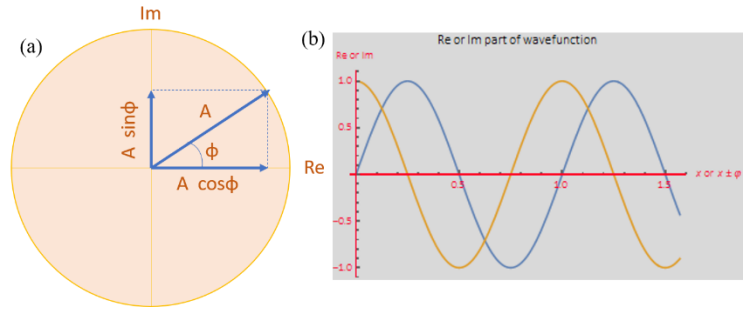

*Fig. S11.* *(a) Complex space describing the complex wave with amplitude A and phase φ, and (b) associated Re and Im parts oscillation with φ and x.*

The complex space is defined with the help of a phasor diagram and it is the $Re$ and $Im$ part of the wave function which oscillate with *x*. The $\varphi$ is the phase shift of the wave from the reference *x* point or real axis in the phasor diagram and $k = 1/\lambda$ is the frequency of the wave. The plot of *Re* and *Im* part of the wavefunction resemble cosine and sine waves, respectively. The square amplitude or intensity of the wavefunction is $= Re^2 + Im^2$ and is constant for all *x*. This is obtained by multiplying the wave function with its complex conjugate and is known as Born rule in Quantum Mechanics [1927 Solvey conference].[17] Thus, any phase shift information in terms of φ is lost while recording intensity which is a measurable quantity. Any information on phase shift needs to be translated into intensity information. However, changing any of the quantity which changes the overall phase angle inside the trigonometric expression, changes the $Re$ and $Im$ part of the wave simultaneously and do not change the overall amplitude, unless an interference experiment is performed with the help of a reference wave. This can be performed through wave interference at an angle (off-axis holography) or along the same axis (in line holography).

Moreover, the amplitude (*A*) should not have any structure in terms of spatial variation in intensity, it only oscillates with change in phase through trigonometric function. Therefore, any amplitude square and multiplication of amplitudes between direct and diffracted waves present alone in any interference formalism should be treated as contributing to uniform background.

However, there are examples where the phase change is brought in different ways to change the overall information, see examples based on Fourier formalism and convolution in real space [section supp. sec. 2.3 and 2.1, 2.2 in main text]. For both WPOA formalism, the change in phase has been considered in terms of change in wavelength ($\lambda$) or wave vector ($k$).[2,12] This is due to potential energy of object alters the wavelength of illuminating electrons through increase in speed when passing through the atom (potential around an atom) and material (mean inner potential). The change in wave vector is also associated with the change in momentum transfer direction (refraction) and finally matches with the wave vector in vacuum to keep the process elastic (see Bloch wave formalism and dispersion surface concept in Ref.[12]). Quantum mechanical Plank's constant ($h$) enters into the formalism through relativistic correction of electron wave vector. This results in a transmission function and only after series approximation within weak phase object approximation (WPOA) the phase information can be obtained in the recorded image [see section S2.6]. In this method, the change in phase becomes proportional to the potential of the object.

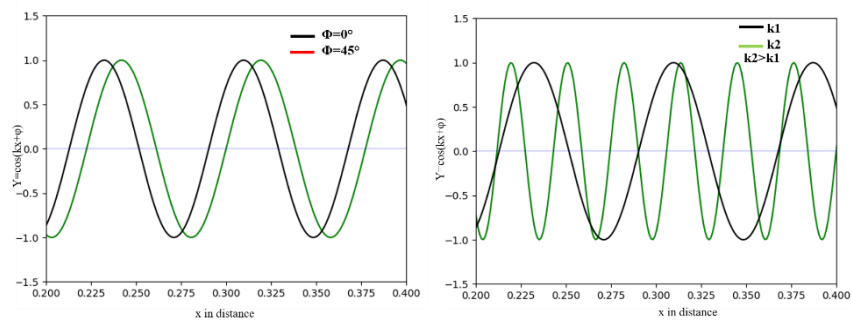

***Fig.S12.*** *Example showing change in plane wave function after phase shift with respect to original wave function when (a) phase shift is added to the original phase within the trigonometric functions, and (b) change in magnitude of wave vector.*

In case of scattering factor-based formalism, it is the interference between the plane wave and scattered spherical wave with amplitude factor given by atom scattering factor for isolated atom. And for both the methods mentioned above, i.e. WPOA and transmitted wave function based on atom scattering factor, the phase change is brought in by series approximation and lens envelop function in diffraction plane which act as a frequency filter and convolution procedure in the image plane.[2,13] However, if we evaluate the cosine and sine part and multiply with amplitude to produce a single number, we mix-up phase -amplitude information in a complicated fashion and recovering phase also becomes difficult. However, if we add the phase shift to the phase of object wave then the square amplitude does not contain the phase information, see the sequence of equations in section S3.

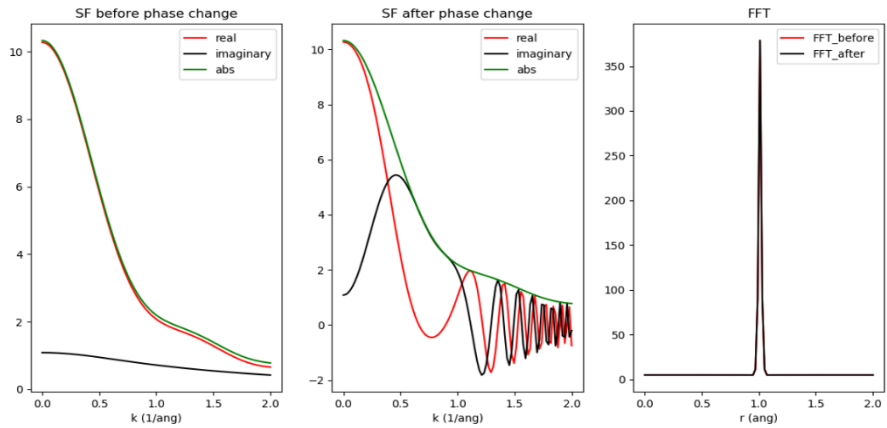

***Fig. S13.*** *The change of image wave function due to aberration of lens (Cs = -35 μm, Δf = 8 nm) when (a) it is added with the phase of image wave and (b) when it is applied after separate evaluation of the exponential terms. (c) Absolute FT*

**Table S1.** *Intensity and FWHM values calculated using Eq. 3 (WPOA) and Eq. 6 (Atom scattering factor) for Mo, S, B, N, Zn and O atoms.*

| Atom | Method | | $I_{max}$ | FWHM |
|---|---|---|---|---|
| Mo (42) | Spence | LR | 22000 | 0.25 |
| | | WLR | 3.5 | 0.06 |
| | Kirkland | both sine &cosine | 52 | 0.5 |
| | | Only sine | 1.05 | 0.75 |
| B (5) | Spence | LR | 3500 | 0.25 |
| | | WLR | 1.06 | 0.06 |
| | Kirkland | both sine &cosine | 2.2 | 0.5 |
| | | Only sine | 1.008 | 0.75 |
| S (16) | Spence | LR | 9800 | 0.25 |
| | | WLR | 2.3 | 0.1 |
| | Kirkland | both sine &cosine | 11 | 0.5 |
| | | Only sine | 1.02 | 0.75 |
| N (7) | Spence | LR | 4600 | 0.25 |
| | | WLR | 1.65 | 0.1 |
| | Kirkland | both sine &cosine | 4.4 | 0.5 |
| | | Only sine | 1.013 | 0.75 |
| Zn (30) | Spence | LR | 15150 | 0.25 |
| | | WLR | 3.23 | 0.1 |
| | Kirkland | both sine &cosine | 70 | 0.45 |
| | | Only sine | 1.06 | 0.63 |
| O (8) | Spence | LR | 5100 | 0.25 |
| | | WLR | 1.7 | 0.1 |

| | Kirkland | both sine &cosine | 4 | 0.45 |
| | | Only sine | 1.01 | 0.63 |

** LR: lense response, WLR: without lens response

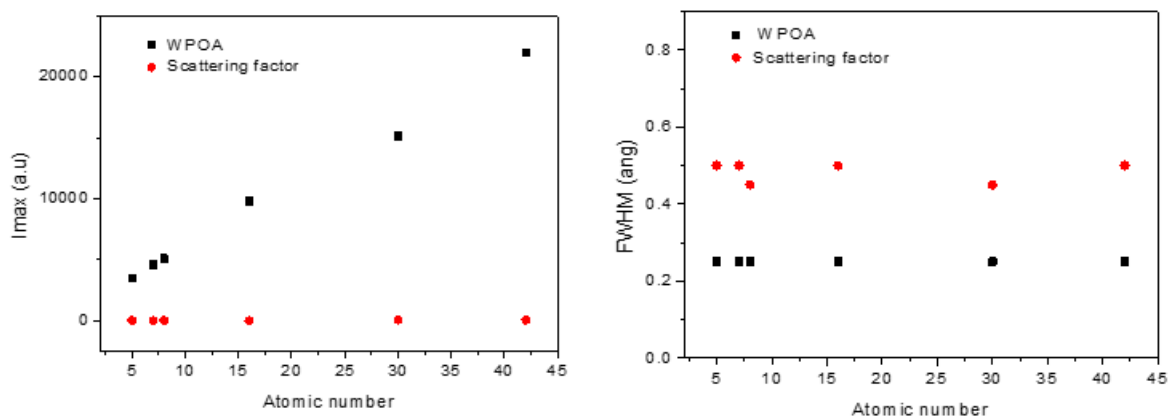

*Fig. S14. Intensity and FWHM values calculated using Eq. 3 (WPOA) and Eq. 6 (Atom scattering factor) for Mo, S, B, N, Zn and O atoms. Values are plotted for only sine part or PCTF function (Cs = -35 μm, Δf = 8 nm) considered for the intensity calculation. For complete ACTF and PCTF consideration based on Eq. 6 see the Table S1 above. The difference in peak intensity and FWHM maximum calculated by two different methods are markedly different.*

## 2.9. Alternative method of image simulation considering atom as electrostatic charge center

This is the alternate method as described in the manuscript section 2.3. The principle is the extension of conventional off-axis electron holography to considering atom as electrostatic charge center which induce interference of wave along complete azimuthal orientations.

The unidirectional hologram intensity equation as given in Eq. S21 is modified for atom charge center by the following expression

$$I(r) = a_1^2 + a_2^2 + 2a_1 a_2 \cos(2\pi q_c r + \Delta\phi) * 2\pi(r_{max} - r)$$

(47)

And the flux balance is given by

$$\int_{-drr}^{drr} I(r)\, dr = \pi(r_2^2 - r_1^2) \times 1 \tag{48}$$

Where, $drr = (r_2 - r_1)/2$ and $r = \sqrt{x^2 + y^2}$

flux balancing between the flux of wave at the plane of atom as given by the coherent rim width $(r_2 - r_1) \times I$, where $I$ is the intensity at a given point within this zone, and resulting interference field given by a square area $r^2 \times I$ (1st law of thermodynamics).

For contribution of various zones in the interference pattern depending on focus see Table S2.

The relationship of phase and resulting magnitude of intensity can be understood as follows. Radial interference geometry modifies the unidirectional straight electron interference fringe to a radial symmetric pattern with maximum intensity at the center of the pattern. Whereas the intensity pattern oscillates periodically with same magnitude in case of unidirectional interference across a mirror plane. Now this phase term appears as carrier frequency that alters the maxima of the radial pattern depending on the radial zone and field strength of that zone. The field strength and corresponding potential information is the object information that can be interpreted.

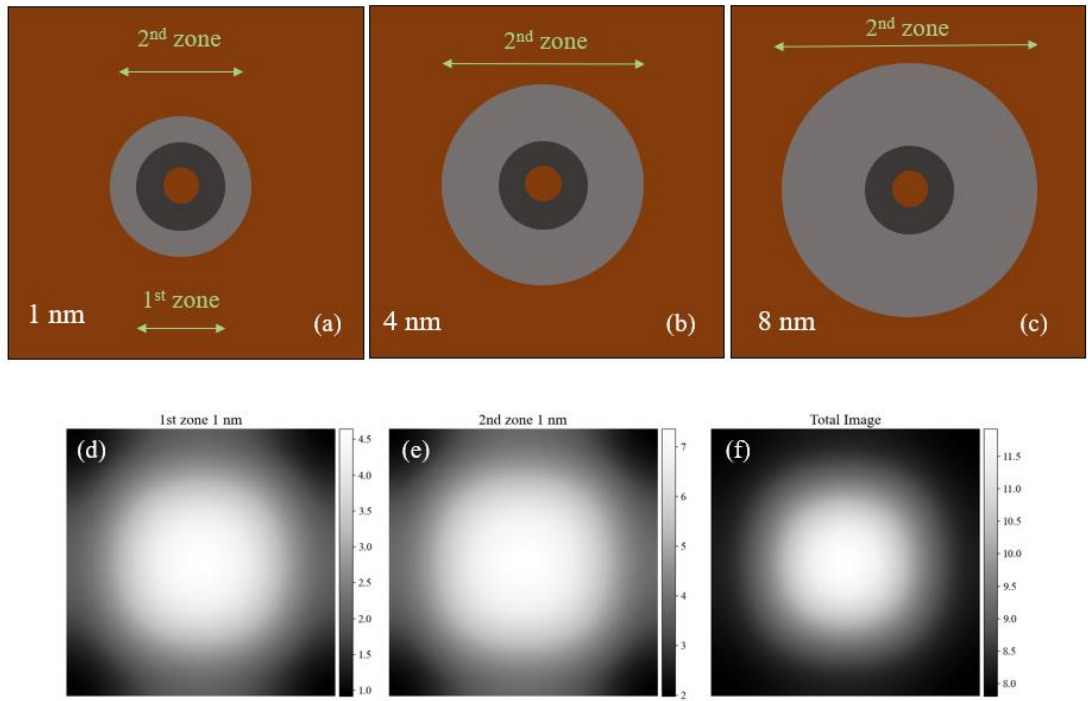

*Fig. S15. potential as a function of distance for isolated B atom, (b) zones divided corresponding to 1 nm focus step with respect to optic axis. Each zone belongs to flat part of the potential (having almost same bending angle). Each zone will contribute to the image contrast incoherently in terms of overlap interference regions depending on focus settings.*

**Table S2.** Various radial zones and corresponding focal length on the optic axis for Mo, S, Zn, O, B, and N atoms for 1, 4 and 8 nm length on the optic axis. Zones are demarcated based on scattering angle falling in the same order of magnitude. The corresponding schematic figure is given in Fig. 8.

For Mo atom (Z=42)

| Radial distance (pm) & zones | Rim width (pm) | Mean scattering angle (rad) | Length on optic axis (nm) | Intensity (with flux balance) |
|---|---|---|---|---|
| 1-1.77 | 0.77 | 0.1558 | 1 | 5.82 |
| 1.77-10.31 | 8.54 | 0.0488 | 1 | 8.45 |
| 10.31-10.70 | 0.39 | 0.0099 | 1 | 16.89 |
| 10.31-16.91 | 6.60 | 0.0070 | 4 | 11.33 |
| 10.31-21.18 | 10.87 | 0.0062 | 8 | 9.56 |

For Zn atom (Z=30)

| Radial distance (pm) & zones | Rim width (pm) | Mean scattering angle (rad) | Length on optic axis (nm) | Intensity (with flux balance) |
|---|---|---|---|---|
| 1-1.38 | 0.38 | 0.1092 | 1 | 6.82 |
| 1.77-8.37 | 6.6 | 0.0275 | 1 | 7.61 |
| 8.76-9.54 | 0.78 | 0.0092 | 1 | 13.73 |
| 8.76-15.36 | 6.60 | 0.0067 | 4 | 9.93 |
| 8.76-18.85 | 10.09 | 0.0061 | 8 | 11.03 |

For B atom (Z=5)

| Radial distance (pm) & zones | Rim width (pm) | Mean scattering angle (rad) | Length on optic axis (nm) | Intensity (with flux balance) |
|---|---|---|---|---|
| 1-2.16 | 1.55 | 0.014 | 1 | 5.09 |
| 2.16-4.49 | 1.64 | 0.0069 | 1 | 7.77 |
| 2.16-7.99 | 6.21 | 0.0052 | 4 | 7.31 |
| 2.16-10.31 | 8.93 | 0.0049 | 8 | 6.76 |

For N atom (Z=7)

| Radial distance (pm) & zones | Rim width (pm) | Mean scattering angle (rad) | Length on optic axis (nm) | Intensity (with flux balance) |
|---|---|---|---|---|
| 1-3.33 | 2.33 | 0.022 | 1 | 8.42 |
| 3.33-5.27 | 1.94 | 0.0067 | 1 | 7.42 |
| 3.33-9.15 | 5.82 | 0.0054 | 4 | 8.41 |
| 3.33-11.87 | 8.54 | 0.0050 | 8 | 8.87 |

For O atom (Z=8)

| Radial distance (pm) & zones | Rim width (pm) | Mean scattering angle (rad) | Length on optic axis (nm) | Intensity (with flux balance) |
|---|---|---|---|---|
| 1-3.32 | 2.32 | 0.0252 | 1 | 8.87 |
| 3.32-5.65 | 2.33 | 0.0075 | 1 | 7.09 |
| 3.32-9.15 | 6.22 | 0.0061 | 4 | 8.40 |
| 3.32-12.25 | 8.93 | 0.0057 | 8 | 8.87 |

For S atom (Z=16)

| Radial distance (pm) & zones | Rim width (pm) | Mean scattering angle (rad) | Length on optic axis (nm) | Intensity (with flux balance) |
|---|---|---|---|---|
| 1-6.04 | 5.04 | 0.043 | 1 | 9.60 |
| 6.04-7.21 | 1.17 | 0.0081 | 1 | 10.58 |
| 6.04-12.25 | 6.21 | 0.0060 | 4 | 11.63 |
| 6.04-15.36 | 9.71 | 0.0055 | 8 | 11.64 |

## 2.10. Effect of Aberration on image contrast

The effect of defocus ($C_1$) and third order spherical aberration coefficient ($C_3$) on the image aberration is given by

$$r_s = MC_3\theta^3 + MC_1\theta \tag{49}$$

Where $r_s$ is the radius of the disk at the Gaussian image point.

Note that spherical aberration and defocus aberration depends on the scattering angle to the third power and linearly, respectively. Opposite values of $C_3$ and $C_1$ reduce the aberration which is a function of scattering angle [Fig. S17(a)]. The effect is to decrease the $r_s$, and improves the resolution of the imaging process. Under best condition (e.g. Scherzer criteria) the optimum balance gives the best minimum value of $r_s$, also called disk of least confusion.

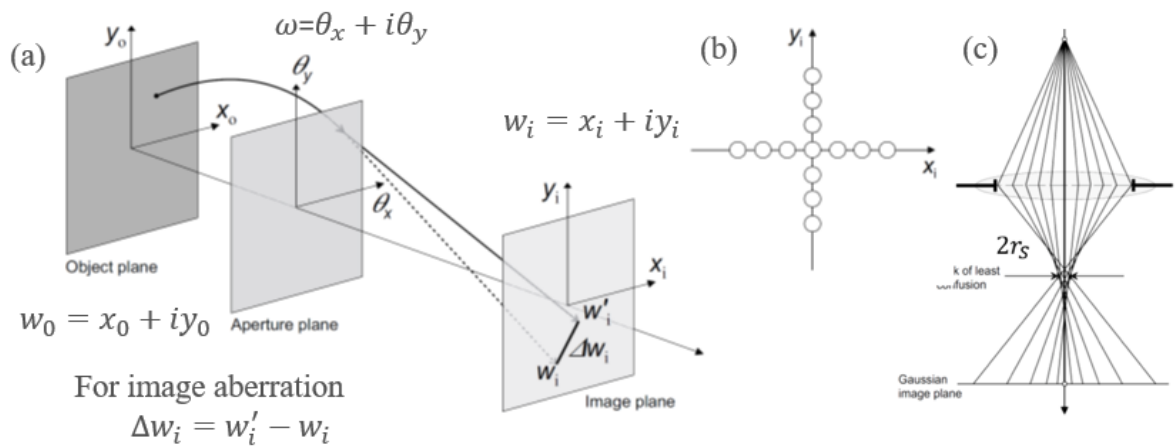

*Fig. S16. (a) Definition of aberration figure in image plane, (b) effect of $C_s$ on object point on image plane, and (c) schematic representation of the effect of positive $\Delta f$ for a given $C_s$ on the image blurring in terms of aberration figure.*[18]

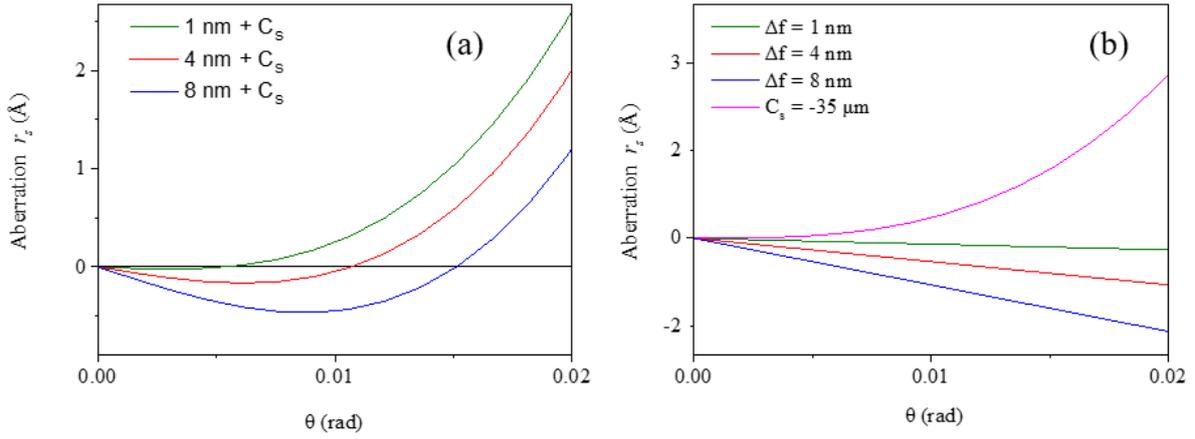

***Fig. S17:*** *(a) Effect of positive $\Delta f$ and negative $C_s$ on $r_s$ for different focus setting of $\Delta f = +1, +4$ and $+8$ nm for $C_3 = -35$ µm. (b) Effect of positive $\Delta f$ and $C_s = 0$ are shown. Note the opposite effect of negative $C_s$ and positive $\Delta f$ on the $r_s$.*

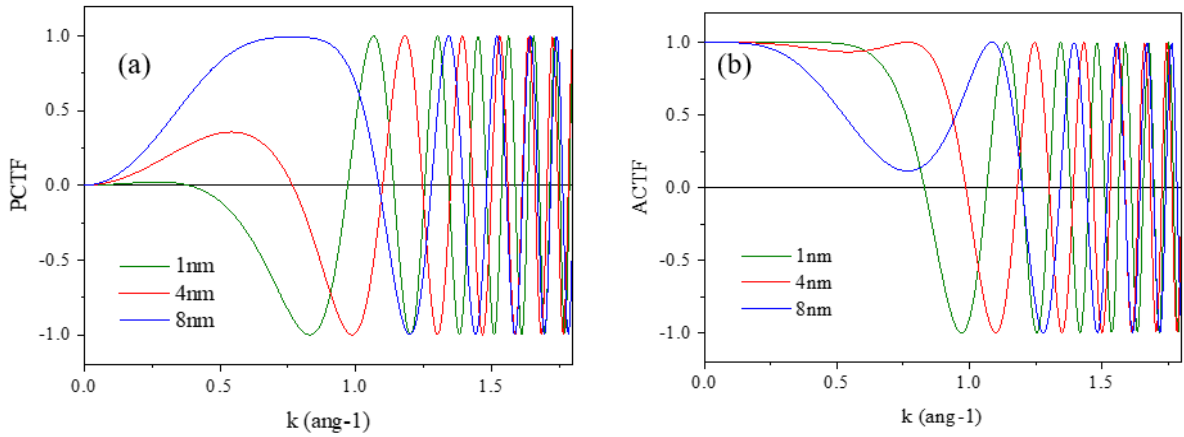

***Fig. S18:*** *(a) PCTF and (b) ACTF function for $C_3 = -35$ µm and potive focus setting of $\Delta f = +1, +4$ and $+8$ nm.*

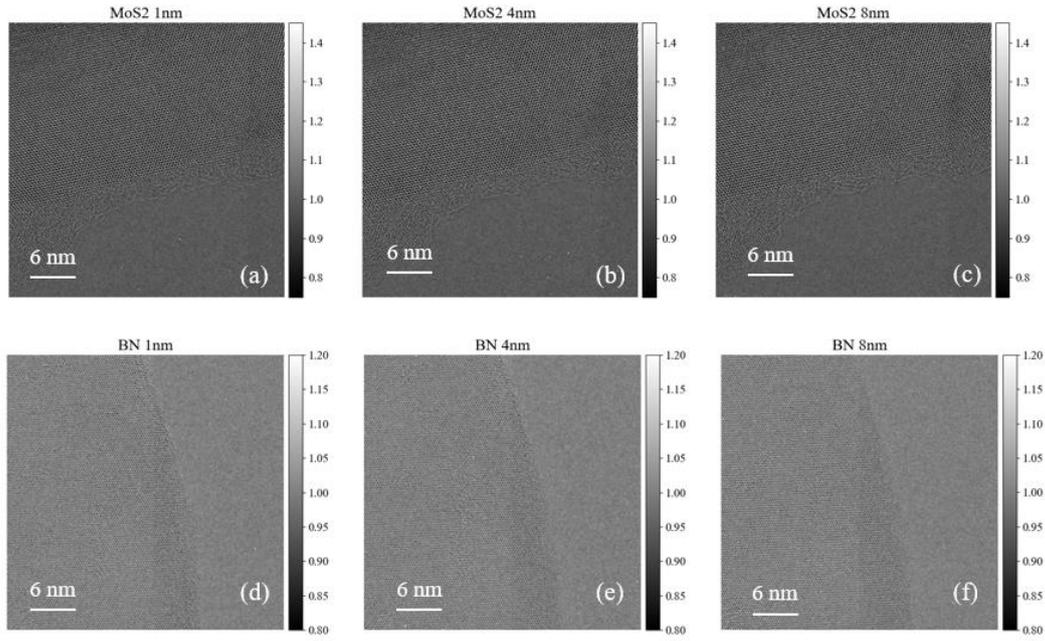

***Fig. S19:*** *Experimental images at three different focus settings of of Δf = +1, +4 and +8 nm for (a)-(c) MoS₂ and (d)-(f) BN.*

**Table S3:** Example scattering angle vs. aberration figure for various combination of $C_s$ and Δf as extracted from Fig. S18.

| Third order spherical aberration $C_s$ (μm) | Defocus Δf (nm) | Scattering angle (rad) | Aberration figure (pm) |
|---|---|---|---|
| −35 | +1 | 0.011 | 31.15 |
| −35 | +4 | 0.011 | 0.67 |
| −35 | +8 | 0.011 | 43.10 |
| 0 | +1 | 0.011 | 10.60 |
| 0 | +4 | 0.011 | 42.42 |
| 0 | +8 | 0.011 | 84.85 |